\documentclass[final]{aipproc}
\layoutstyle{6x9}

\newcommand{\bea}{\begin{eqnarray}}
\newcommand{\eea}{\end{eqnarray}}
\newcommand{\be}{\begin{equation}}
\newcommand{\ee}{\end{equation}}

\def\spose#1{\hbox to 0pt{#1\hss}}

\newcommand{\bx}{\ensuremath{\mathbf{x}}}
\newcommand{\br}{\ensuremath{\mathbf{r}}}
\newcommand{\bk}{\ensuremath{\mathbf{k}}}
\newcommand{\bF}{\ensuremath{\mathbf{F}}}
\newcommand{\bR}{\ensuremath{\mathbf{R}}}
\newcommand{\bu}{\ensuremath{\mathbf{u}}}
\newcommand{\bn}{\ensuremath{\mathbf{n}}}

\newcommand{\bff}{\ensuremath{\mathbf{f}}}
\newcommand{\tbu}{\ensuremath{\tilde{\mathbf{u}}}}

\newcommand{\ve}[1]{\ensuremath{\mathbf{#1}}}
\newcommand{\D}[1][ ]{\ensuremath{\mathrm{d}^{#1} }}

\begin{document}
 
\title{Infinite self-gravitating systems \\
and cosmological structure formation}
\author{Michael Joyce}{address={
Laboratoire de Physique Nucl\'eaire et de Hautes Energies, UMR-7585 \\
Universit\'e Pierre et Marie Curie --- Paris 6 \\
75252 Paris Cedex 05, France}}

\begin{abstract}
 The usual thermodynamic limit for systems of classical self-gravitating 
point particles becomes well defined, as a {\it dynamical} problem, using
a simple physical prescription for the calculation of the force,
equivalent to the so-called ``Jeans' swindle''. The relation of the 
resulting intrinsically out of equilibrium problem, of particles
evolving from prescribed uniform initial conditions in an infinite 
space, to the one studied in current cosmological models (in an
expanding universe) is explained. We then describe results of a
numerical study of the dynamical evolution of such a system, starting 
from a simple class of infinite ``shuffled lattice'' initial
conditions. The clustering, which develops in time starting from scales around
the grid scale, is qualitatively very similar to that seen 
in cosmological simulations, which begin from lattices with applied 
correlated displacements and incorporate an expanding spatial background. 
From very soon after the formation of the first non-linear structures, a 
spatio-temporal scaling relation describes well the evolution of the 
two-point correlations. At larger times the dynamics of these correlations 
converges to what is termed ``self-similar'' evolution in cosmology,
in which the time dependence in the scaling relation is specified entirely by
that of the linearized fluid theory. We show how this statistical
mechanical ``toy model'' can be useful in addressing various questions 
about these systems which are relevant in cosmology. Some of these 
questions are closely analagous to those currently studied in the 
literature on long range interactions, notably the relation of the
evolution of the particle system to that in the Vlasov limit 
and the nature of approximately quasi-stationary states.
\end{abstract}
 
\maketitle
 
\section{Introduction}

The evolution of systems of a very large number of classical point 
particles interacting solely by Newtonian gravity is a paradigmatic
problem for the statistical physics of long range interacting systems. 
It is also a very relevant limit for real physical problems studied 
in astrophysics and cosmology, ranging from the formation of galaxies
to the evolution of the largest scale structures in the universe.
Indeed current theories of the universe postulate that most of the
particle-like clustering matter in the universe is ``cold'' 
and ``dark'', i.e. very non-relativistic and interacting
essentially only by gravity, which means that the purely
Newtonian approximation is very good one over a very great range of
temporal and spatial scales. In practice, as we discuss further below,
the very large numerical simulations used to follow the evolution of
the structures in the universe, starting from the initial conditions
on the density perturbations inferred from measurements of the 
fluctuations in the cosmic microwave background, are completely 
Newtonian. While simulations in this context have developed
impressively in size and sophistication over the last three 
decades,  the results they provide remain essentially 
phenomenological in the sense that our analytical understanding 
of them is very limited.

Our first aim in this contribution is to clarify, in a statistical
physics language, precisely what the relevant problem is which is 
currently studied in the context of the problem of structure formation
in cosmology. The essential point is that it corresponds simply
to a specific infinite volume limit of the Newtonian problem. 
A well defined limit of the resulting equations of motion is given 
by the case that the universe does not expand. This case, we explain,
corresponds to the most natural definition of the usual thermodynamic 
limit of the Newtonian problem. We note that this limit is distinct 
from the ``mean-field'' or ``dilute'' one sometimes 
considered\footnote{See contribution of P.H. Chavanis.} 
in the literature, and in which, under certain circumstances, one
can define thermodynamic equilibria. In the limit we consider the
system has no thermodynamic equilibrium, and the system is always
intrinsically out of equilibrium.

The main body of this contribution will then be given over to the
study of this (usual) thermodynamic limit of gravity, for a particular
set of initial conditions. We will describe and characterise the
evolution of clustering as observed in a set of numerical 
simulations, summarising the results of a series of recent publications
with our collaborators \cite{sl1, sl2, sl3}. Further we will highlight
the fact that this evolution is qualitatively very similar to that seen in 
cosmological simulations (in an expanding universe). It is characterised
notably by the fact that it is ``hierarchical'', which means that the
non-linear structures form at a continually increasing scale,
propagating towards larger scales at a rate which is fixed by the 
spectrum of initial fluctuations in the initial
conditions. The evolution is observed also to tend
asymptotically to what cosmologists call a ``self-similar'' behaviour,
which means that the correlations manifest a simple spatio-temporal
scaling behaviour. While the temporal part of this evolution may
be explained by a simple linear fluid model for the growth of 
perturbations, the form of the spatial part remains, as in the
cosmological model, unpredicted. 

We will then outline a model we have developed which describes 
very well the evolution of our simulations at early times. It 
is a two phase model, linking a perturbative treatment of the 
evolution at early times to an approximation in which the force
on particles is dominated essentially by their nearest neighbour.

In the last part we will turn to the relevance
of our analysis to open unresolved questions in the cosmological
problem. Indeed one of our motivations has been that we may
be able to use our system, which is relatively simple compared to 
a cosmological simulation, in this way as a kind of ``toy model''. 
We focus here specifically on the ``problem of discreteness'' in 
cosmological simulations: these simulations simulate a number
of particles which is many orders of magnitude less than that in
the physical problem, and the question is how this limits the
resolution of the corresponding results. Stated more formally,
the desired physical limit is the appropriate Vlasov-Poisson 
system, and the problem is to understand the finite $N$ 
effects\footnote{We will explain below that in the infinite 
volume limit it is, more precisely, a question of the relation
between the evolution of the particle system, with finite 
particle {\it density}, and that in the Vlasov limit.}. We 
discuss what we learn about this problem from
the case of the shuffled lattice. 

In our conclusions we suggest some directions for future work.
In particular we discuss briefly the interest of relating out
results to now very popular phenomenological models for the
evolution of cosmological simulations, so-called ``halo models''.
The central objects in this description (halos) have, apparently,
very similar properties to those observed to be formed through
the violent relaxation of finite self-gravitating systems. We
note that the nature of these states of the finite system is
analogous to the quasi-stationary states observed and much
studied in one dimensional toy 
models\footnote{See the contribution of S. Ruffo in this volume.}. 

\section{Newtonian gravity in the infinite volume limit}

Let us start from a system of $N$ identical point particles
interacting by purely Newtonian gravity, i.e., with equations of
motion 
\be
\label{Nbody-eom}
\ddot\br_i=
-{Gm}\sum_{j\ne i}\frac{\br_i-\br_j}{|\br_i-\br_j|^3}
\,
\ee
where dots denote derivatives with respect to time, $m$ is the mass of
the particles, $G$ is Newton's constant, and $\br_i$ is the position 
of the $i$-th particle. If $N$ is finite the evolution is well defined
\footnote{This is true of course assuming that $\br_i \neq \br_j$ for 
all $i\neq j$. Starting from initial conditions in which this is true
--- as we always shall here --- this undefinedness of the problem 
associated with this singularity of the force is never relevant, in 
a finite time. Whether this singularity is ``left naked'' or removed
by a smoothing at some scale is, however, of course relevant to the
very long time behaviour of the system. Indeed the so-called ``gravothermal
catastrophe'' in these systems arises from the divergence of the 
microcanonical entropy due to configurations in which pairs of particles 
approach one another arbitrarily close, allowing other particles in the
system to explore an unlimited phase space (in momentum). We will not
explore the question here of the behaviour of these systems in
the limit that $t \rightarrow \infty$. See further comments below when
we discuss the use of a regularisation at small scales of the force
in numerical simulations.}, whether the
system is open (i.e. in an infinite space), or enclosed in a region
(with some appropriate boundary conditions). 

Let us 
consider now the usual thermodynamic limit, i.e., 
\be
\label{TDlimit}
N \rightarrow \infty, \, V \rightarrow \infty, \,\, {\rm at}
\,\, n_0=\frac{N}{V}={\rm constant}
\ee
For example we can consider taking $N$ points which we randomly 
distribute in the volume $V$, increasing the number $N$ 
in strict proportionality with the volume $V$. A little consideration
of Eq.~(\ref{Nbody-eom}) shows that there is a problem when we
take the limit: the sum on the right hand side, giving the gravitational
acceleration of any given particle, is not unambiguously defined 
by this limiting procedure.  To see this more explicitly, let
us suppose that we choose our volume to be a spherical volume  $V(R_s, \br_0)$ 
centred on some arbitrary fixed point, at vector position $\br_O$,
and of radius $R_s$, and that we take the infinite volume limit
by increasing this radius $R_s \rightarrow \infty$. The force 
per unit mass for the point $\br_i$ is then
\be
\label{force1}
\bF (\br_i) = -{Gm} \lim_{R_s \rightarrow \infty}
\sum_{j\ne i, j \in V(R_s, \br_0)} \frac{\br_i-\br_j}{|\br_i-\br_j|^3}
\,.
\ee
Defining $n_i (\br) = \sum_{j \neq i} \delta (\br - \br_j)$,
and $\delta n_i (\br) = \sum_{j \neq i} \delta (\br - \br_j) -n_0$
[where $\delta(\br)$ is the Dirac delta function], this can be 
rewritten as 
\bea
\label{force2}
\bF (\br_i) &=& -{Gm} \lim_{R_s \rightarrow \infty}
\int_{ V(R_s, \br_0)} d^3 x n_i(\br) \frac{\br_i-\br}{|\br_i-\br|^3}
 \nonumber\\
 &=& -{Gm} \lim_{R_s \rightarrow \infty}
\left[n_0 \int_{ V(R_s, \br_0)} \frac{\br_i-\br}{|\br_i-\br|^3}
+\int_{ V(R_s, \br_0)} d^3 r \delta n_i(\br) \frac{\br_i-\br}{|\br_i-\br|^3}
\right] \,
\eea

\subsection{Force due to mean density}

The first term inside the brackets, representing the force per unit mass
on the particle due to the mean density $n_0$ on a particle inside the sphere, 
gives, using Gauss' theorem, for any point $\br_i \in V(R_s, \br_O)$
\be
\label{force-spherical}
\bF_0 (\br_i)= -\frac{4 \pi G n_0 m}{3} \left( \br_i -\br_0 \right)\,.
\ee
This result is independent of $R_s$, and thus is also the expression
of this contribution to the force in the limit that $R_s \rightarrow
\infty$. We see, however, that it depends explicitly on the choice of 
origin. We will analyse the second term in the square brackets
in Eq.~(\ref{force2}), due 
to the correction to uniformity, in the next subsection. We will
verify that this latter contribution may converge to a well defined value, 
independently of how the limit is taken, and thus that it 
cannot remove this dependence of the full force on the chosen point.

To make the problem of Newtonian gravity in the usual thermodynamic
limit well defined we need a uniquely defined force acting on particles.
Given the arbitrariness in the result of the above calculation, we must
thus give a prescription for calculating this force. As the ambiguity 
arises from the uniform component of the particle density, an evident 
choice for such a prescription is one which makes the contribution of 
the mean density zero for any particle. This is attained by any prescription 
for the sum  in which the force on any given particle is determined by summing
in a {\it symmetric} manner about that particle, e.g., 
\bea
\label{symmetric-force-defn}
\bF (\br_i) &=& -{Gm} \lim_{R_s \rightarrow \infty}
\sum_{j\ne i, j \in V(R_s, \br_i)} \frac{\br_i-\br_j}{|\br_i-\br_j|^3} \\
&=& -{Gm} \lim_{R_s \rightarrow \infty}
\left[n_0 \int_{ V(R_s, \br_i)} \frac{\br_i-\br}{|\br_i-\br|^3}
+\int_{ V(R_s, \br_i)} d^3 r \delta n_i(\br) \frac{\br_i-\br}{|\br_i-\br|^3}
\right] \\
\label{symmetric-force-deltan}
&=& -{Gm} \lim_{R_s \rightarrow \infty}
\left[\int_{ V(R_s, \br_i)} d^3 r \delta n_i(\br) \frac{\br_i-\br}{|\br_i-\br|^3}
\right] \,.
\eea
where $V(R_s, \br_i)$ is the spherical volume of radius $R_s$
{\it centred on the $i$-th point}. The force on a particle is then
that due only to the fluctuations around the mean density. 

This prescription for the calculation of the force is essentially
equivalent to what is known as the ``Jeans' swindle'' \cite{jeans, binney}.
To treat the dynamics of perturbations to an infinite self-gravitating 
pressureful fluid, Jeans ``swindled'' by perturbing about the state of 
constant mass density and zero velocity, which is not in fact a solution 
of the fluid equations he analysed. Formally the ``swindle'' involves 
removing the mean density from the Poisson equation, so that the 
gravitational potential is sourced only by fluctuations to this mean 
density. This is evidently equivalent to the prescription we have just
given for the gravitational force, as seen explicitly 
in Eq.~(\ref{symmetric-force-deltan}). 

A very nice discussion of the mathematical basis of this ``Jeans'
swindle'' is given by Kiessling\cite{kiessling}, who emphasises
that the presentation of what Jeans did as a ``mathematical swindle''
is misleading. The ``swindle'' can in fact be given a perfectly firm
mathematical basis when it is understood, as we have presented
it above, as a physical regularisation of the {\it force}. 
The mathematical inconsistency in Jeans' analysis arises because 
it is done in terms of potentials which are badly defined, even
in the thermodynamic limit as we have defined it. 
In Ref.~\cite{kiessling} Kiessling shows that the Jeans swindle
corresponds to a regularisation of the force given by the prescription
used above in Eq.~(\ref{symmetric-force-defn}), which just corresponds
to the physical prescription that the force on all particles be zero 
for an infinite constant density. Kiessling notes that an equivalent
more physical form of the prescription is 
\be
\label{symmetric-force-defn-screening}
\bF (\br_i) = -{Gm} \lim_{\mu \rightarrow 0}
\sum_{j\ne i} \frac{\br_i-\br_j}{|\br_i-\br_j|^3} e^{-\mu|\br_i-\br_j|}
\ee
where the sum now extends over the infinite space\footnote{This means
that the thermodynamic limit has been taken before the limit in $\mu$.},
i.e., the value of the gravitational force in the thermodynamic limit 
is given as the limit of a screened gravitational force. One may thus 
actually consider the dynamics we describe below as characterising that 
of a system of particles with a screened gravitational force, but where 
the screening scale ($=\mu^{-1}$) is much larger than the length scales
on which ``non-linear'' clustering develops (see below). 

\subsection{Force due to fluctuations}

Having given the prescription which makes the force zero in the 
usual thermodynamic limit for the case of an exactly uniform
distribution of matter, we must consider also whether this prescription
makes the force well defined in a distribution of particles, in which
there are necessarily deviations from perfect uniformity, due to the
discrete nature of the particles. To answer this question we must
evidently consider {\it how} these particles are distributed in the infinite 
space, i.e., the nature of the fluctuations which they give rise to.
For any distribution which one proposes to study as initial conditions
of the infinite self-gravitating, one must ensure that the force
is then well defined. 

To prescribe (and describe) how particles are distributed in infinite 
space we use the language of stochastic point processes, i.e., we
consider the points to be distributed by a stochastic process, with
certain properties. We assume only that these processes have only 
the following properties (which we require to employ them usefully here):
\begin{itemize}
\item {\it spatial ergodicity}, i.e., the value of relevant quantities 
charactering their fluctuations/correlations in the infinite volume limit
 may be recovered by taking the ensemble average.  
\item {\it statistical homogeneity}, i.e., averages are
invariant under any rigid translation of the system.
\item {\it uniformity}, i.e., they have a well defined {\it non-zero}
      mean density, $n_0 =\lim_{N,V \rightarrow \infty} (N/V)= \langle n(\bx) \rangle$
(where $\langle ... \rangle$ denotes the ensemble average).
\end{itemize}

In this framework the force on a particle, and also the force
field (per unit mass) at any point in space, are stochastic variables. 
Let us consider the latter quantity\footnote{We analyse here, for
simplicity, the {\it unconditioned} force field, i.e., the force per
unit mass at an arbitrary spatial point, rather than the force on an actual 
point of the distribution. These quantities will in general be 
different, notably for the small scale properties related to divergences
of the force at zero separation. For the study of convergence/divergence
of the force due to the long-range nature of the force, the quantities
are equivalent.}, denoting 
it $\bff (\br)$. It may
be written as an integral over the stochastic variable $n(\br)$ as
\be
\label{symmetric-force-defn-stoch}
\bff ( \br )= -{Gm} \lim_{R_s \rightarrow \infty}
\left[\int_{ V(R_s, \br )} d^3r^\prime n (\br^\prime) \frac{\br-\br^\prime}{|\br-\br^\prime|^3}
\right] \,.
\ee
The question of the well-definedness of the force may then be phrased as
follows: is the probability distribution function (PDF) of the 
variable $\bff$, defined at any finite $R_s$, well defined in the 
limit $R_s \rightarrow \infty$? An answer may be given as follows.
Using the convolution theorem for Fourier transforms (FT) we can infer
that\footnote{To do this FT, which is valid only in the sense
of distributions, it is more appropriate to write the thermodynamic
limit as in Eq.~(\ref{symmetric-force-defn-screening}).} 
\be
|\tilde{\bff}^2 (\bk )| \propto  k^{-2} |\tilde{\delta n} (\bk )|^2
\ee
where the tilde denote the FT (and we have used the fact that 
the FT of $\bx/|\bx|^3$ is proportional to $\bk/|\bk|^2$). The square of 
the variance of any stochastic variable is, up to an appropriate 
normalisation in the volume\footnote{The exact definition will be given
in the next section.}, its power spectrum (or structure factor). 
The integral of this quantity, it can be shown easily (see 
e.g. \cite{book}) is proportional to the one point 
variance of the variable, from which is follows here 
that 
\be
\langle {\bff}^2 \rangle \propto \int d^3 k \, 
k^{-2} P(\bk)
\ee
where $P(k)$ is the power spectrum for the density field $n(\br)$
of the point process. If this variance of the force is finite, the 
force PDF is necessarily well defined. 
This means that a {\it sufficient} condition for the force to
be defined in the usual thermodynamic limit is just that
$k^{-2} P (\bk)$ be integrable. Since we are interested here only 
in the possible divergences in all these quantities due to the 
long distance behaviour of the 
fluctuations\footnote{Note that we have already implicitly made this assumption
in writing Eq.~\ref{symmetric-force-defn-stoch}, as we have ignored 
the divergences when $\bx$ falls on an occupied point if there is 
no regularisation of the force at zero separation.}, it is 
therefore sufficient that we require 
$\lim_{k\rightarrow 0} k P (\bk)=0$. Thus it follows that the force is
well defined in the infinite volume limit {\it if $P(\bk)$
diverges more slowly at small $k$ than $k^{-1}$}. 
This can be compared with the condition of uniformity of the point 
process, which requires
that $P(\bk)$ be integrable, and thus only that it not diverge
faster than $k^{-3}$ as $k \rightarrow 0$. We can conclude that 
the force PDF is, at least, well defined for this sub-class of all
uniform point processes.

A Poissonian point process (i.e. particles thrown randomly in the
infinite volume with any finite mean density $n_0$), which corresponds
to a constant $P(\bk)$, should thus give a well defined force in the 
thermodynamic limit. The PDF of the force on a particle for this case 
(with the same implicit regularisation we have described) was first
calculated by Chandrasekhar in the forties\cite{chandra43}, and 
gives the so-called Holtzmark distribution (see \cite{book} for 
the full expression and a simplified derivation). The PDF is
indeed well defined, and thus the dynamical problem in
the thermodynamic limit with the given prescription. The
variance is, in fact, infinite due to the small scale behaviour 
of the force: the probability at very large forces decays 
slowly (as a power-law) due to the unbounded growth of the 
interparticle force in configurations
in which two particles are arbitrarily close. Imposing a cut-off
regulating the divergence in the force at small separations,
it is simple to determine (see \cite{book}) how the variance 
of the force, which
is then finite, diverges as this cut-off goes to zero. We
will consider in detail below the case 
of ``shuffled lattice'' initial conditions. A detailed
treatment of the statistics of the force field in these
distributions is given in Ref.~\cite{gabrielli_06}.
The power spectrum $P(k)$ in this case (see below)
is proportional to $k^2$ at small $k$, and it may be
shown explicitly that the variance of the force is 
therefore finite in the defined thermodynamic limit, 
modulo divergences due to the small scale behaviour 
of the force. Indeed it is shown in Ref.~\cite{gabrielli_06}
that the exact gravitational force PDF decays exponentially 
at large values of the field in distributions of this
type in which there is zero probability of particle overlap, but 
as in the Poisson case (with infinite variance) if there 
is a non-zero probability for such configurations\footnote{An
analogous treatment of generic power law interactions can
be found in Ref.~\cite{andrea_1dforces}.}.

\subsection{Relation to expanding universe case}

The principal conclusion of the preceding two sections is that 
there is a natural thermodynamic limit for the dynamical problem 
of self-gravitating particles in the Newtonian limit. The equations
of motion of the particles are simply Eq.~(\ref{Nbody-eom}), where the
sum is now prescribed to be taken symmetrically about each 
particle\footnote{In practice this sum will be performed in the 
numerical simulations described below using the Ewald sum method. 
In this technique (see e.g. Ref.~\cite{hernquist}) the 
long distance part of the sum is converted to a sum in reciprocal 
space, from which the badly defined term at $\bk=0$, due to the mean
density, is removed.}. Provided the distribution of particles considered 
has fluctuations which decay sufficiently rapidly at large 
distances, this prescription, we have seen, gives a force on particles
which is well defined\footnote{The power spectrum $P(\bk)$ is the FT of the
reduced two point density-density correlation function, and thus
the constraints on the sufficiently slow divergence at small $k$ 
of the former correspond to constraints on the rapidity of decay
of the latter at large separations. Alternatively one can also
relate $P(\bk)$ to the variance of mass in a volume, the latter being
given by the integral of the former weighted by the FT of the 
window function describing the volume (see e.g. Ref.~\cite{book}). 
For  $P(k\rightarrow 0) \sim k^n$, with $n<1$, this gives, for example,
that the variance in particle number in spheres of radius decays
as $R^{3-n}$. We have therefore shown above that the gravitational
force has a finite variance due to fluctuations at large distances
if the variance of these fluctuations grows more slowly with
$R$ than $R^4$.}.  
Further, as we will see further below,
the behaviour of fluctuations at arbitrarily large scales is
a property of the distribution which is not modified by the
gravitational evolution, so the force will then be defined
at all times if it is defined in the initial distribution\footnote{
The linearized fluid theory (see below) describes the evolution 
fluctuations at large scales, giving a simple $k$-independent 
evolution for the power spectrum at small $k$. The latter thus
maintains its initial functional form.}.

The simplest way to understand the relation of this dynamical
problem to that treated in cosmology is by comparing these equations
of motion to those which apply in numerical simulations of structure
formation in the universe. These are written (see
e.g. Ref.~\cite{efstathiou_init})
\be
\label{Nbody-eom-cosmo}
\ddot\bx_i + 2 \frac{\dot a(t)}{a(t)}\dot \bx_i=
-\frac{Gm}{a^3}\sum_{j\ne i}\frac{\bx_i-\bx_j}{|\bx_i-\bx_j|^3}
\,
\ee
where $\bx_i$ are the positions (in ``comoving coordinates'') of the 
particles, and $a(t)$ is the expansion rate of the cosmological
model. The sum on the right-hand side, defining the force per unit
mass on the particles, is evaluated in precisely the way we have
just discussed, i.e., symmetrically, so that the force on each
point is zero when we neglect the density {\it fluctuations} 
associated with the points\footnote{In practice, just as in the
simulations we will describe below, the sum is normally implemented
numerically using an Ewald summation method.}. The thermodynamic
limit we have described for the purely Newtonian problem simply
corresponds to ${\dot a} (t)=0$, i.e. to the formally
non-expanding limit of the equations. 

The equations of motion Eq.~(\ref{Nbody-eom-cosmo}) for the cosmological
case, where ${\dot a} (t) \neq 0$ and $a(t)$ obeys the correct equation
derived from the full general relativistic equations, can in fact
be derived also in a purely Newtonian framework, but using a limit
different to the one we have considered. To see this let us return  
to Eq.~(\ref{Nbody-eom}), but with the sum now 
evaluated as given in Eq.~(\ref{force1}), i.e. choosing a specific
centre and summing the force in spheres centred on it. Taking now 
only the contribution to the force due to the mean 
density, i.e., neglecting the effect of deviations from
uniformity, Eq.~(\ref{force-spherical}) gives
the equation of motion for any particle
\be
\label{eom-uniform}
{\ddot \br_i}= -\frac{4 \pi G n_0 m}{3} \left( \br_i -\br_0 \right)\,.
\ee
To derive this formula we only used the fact that $n_0$ is well defined
at any given instant, but did not assume that it was independent of time.
It follows that we may seek, consistent with our assumptions, a solution 
of these equations describing a simple global dilatation of the 
whole (spherical) system, i.e. of the form
\be
{\br_i}(t) - {\br_0}=R(t) \left[{\br_i}(t_0) - {\br_0}\right]
\ee 
where $R(t_0)=1$. In this case $R(t)$ must then satisfy the equation
\be
\label{frw-1}
\frac{\ddot{R}}{R}=-\frac{4\pi G}{3} \rho (t)= -\frac{4\pi G}{3} 
\frac{\rho_0}{R^3} \,,
\ee
where we have defined $\rho(t)=mn_0(t)$, the mean mass density at time
$t$, and $\rho_0=\rho (t_0)=mn_0(t_0)$. This equation may be integrated
directly to give
\be
\left(\frac{\dot{R}}{R}\right)^2
=\frac{8\pi G}{3}\left[\frac{\rho_0}{R^3}+\frac{\kappa}{R^2}\right] 
\ee
where $\kappa$ is a constant of the motion, proportional to the conserved
``energy'':  
\be
\frac{8\pi G}{3} \kappa=  \frac{1}{2} \dot{R}^2 - \frac{GM_0}{R} 
\,.
\ee
where $M_0$ is a mass density proportional to the (constant) mass 
inside the sphere defined by any particle and the origin of the
coordinates. The dynamics of the dilatation is clearly equivalent
to that of a single particle in a central gravitational field, with
$\kappa=0$ determining the critical escape velocity at the initial 
time.

These equations are identical to the Friedmann equations, derived within 
the framework of general relativity (see e.g. \cite{peebles}) and
which describe an expanding (or contracting) universe, for the case 
that the expansion (or contraction) is sourced only by non-relativistic
matter. In practice all current standard-type cosmological models 
have in common that the universe is described very well by these
equations (i.e. in which the observed expansion of the universe 
is driven only by non-relativistic matter\footnote{We will remark below 
on the addition of a ``cosmological constant'', which is now believed
to cause an acceleration of the observed expansion at recent epochs.}) 
in the epoch at which structure formation develops.
The constant  $R(t)$ and $\kappa$  acquire a different
physical meaning to that ascribed to them in the Newtonian derivation
--- the former describes stretching in time of the spatial sections of the 
four dimensional space-time and the latter the curvature of these
spatial sections --- but this difference in interpretation is of no
relevance in the treatment of the problem of structure formation.

To recover the full equations  used in this 
context, i.e. Eq.~(\ref{Nbody-eom-cosmo}), it suffices now to change
coordinates, defining the comoving coordinates of any particle by
\be
\br_i (t) - \br_0 = R(t) \bx_i (t)
\ee
Substituted in Eq.~(\ref{Nbody-eom}) this gives 
\bea
\ddot\bx_i +\frac{\dot R}{R} \dot\bx_i &=&
-\lim_{R_s \rightarrow \infty} \frac{Gm}{R^3} \sum_{j\ne i, j \in V(R_s,
0)} \frac{\bx_i-\bx_j}{|\bx_i-\bx_j|^3} -\frac {\ddot R}{R} \bx_i 
\label{Nbody-eom-change-1} \\
&=& -\frac{Gm}{R^3} 
\left[\lim_{R_s \rightarrow \infty}  \sum_{j\ne i, j \in V(R_s,0)} 
\frac{\bx_i-\bx_j}{|\bx_i-\bx_j|^3} -\frac {4\pi}{3} n_0 \bx_i \right]
\label{Nbody-eom-change-2} \\
&=& -\frac{Gm}{R^3} 
\lim_{R_s \rightarrow \infty}  \sum_{j\ne i, j \in V(R_s, \bx_i)} 
\frac{\bx_i-\bx_j}{|\bx_i-\bx_j|^3} \,. 
\label{Nbody-eom-change-3}
\eea
where here the sum over $j \in V(R_s, \bx)$ is over all the particles 
in the comoving sphere of radius $R_s$ (i.e. over the physical sphere of 
radius $R(t)R_s$ at time $t$) centred at the comoving coordinate 
position $\bx$. Making the identification of 
$R(t)$ with $a(t)$, this last form is 
precisely Eq.~(\ref{Nbody-eom-cosmo}), with the prescription 
for the summation of the force made explicit. Note that we have used 
Eq.~(\ref{frw-1}) in the passage between Eq.~(\ref{Nbody-eom-change-1}) 
and Eq.~(\ref{Nbody-eom-change-2}), and in passing 
from Eq.~(\ref{Nbody-eom-change-2}) to 
Eq.~(\ref{Nbody-eom-change-3}) we have assumed that the 
force due to the fluctuations (i.e. once the term due to the 
mean density is subtracted) is well defined independently of
how the sum is taken. The criterion for this are identical
to those discussed above.

The only missing element in this derivation with respect to
the full cosmological simulations currently studied is 
the co-called ``cosmological constant''. This enters only
as a modification of the Friedmann equation and thus of
the function $a(t)$ in the Eq.~(\ref{Nbody-eom-cosmo}).
We note that it in fact can be derived in a completely
Newtonian framework, by generalising the above treatment
with an additional classical repulsive force acting 
between all pairs of particles and linearly proportional 
to their vector separation. Because of its linearity it plays no role
in the evolution of fluctuations, other than that encoded
in the modification it causes of the uniform expansion. 

In summary the equations considered in cosmology to model
formation of structure in the universe can be understood
in a purely Newtonian framework as those describing the 
dynamics of particles, enclosed in a very large sphere
and initially almost uniformly distributed with velocities 
close to those described by a homogeneous dilatation 
of this sphere (modelling the Hubble expansion). Formally
we can consider the limit of the corresponding equations
of motion in which there is no expansion. Physically this
limit corresponds not to a very large static sphere --- 
there is no static solution for such a configuration ---
but to a different regularisation of the gravitational
force, which we have explained is naturally considered
as the usual thermodynamic limit of the Newtonian problem
of particles distributed uniformly in an infinite space.

In this precise sense one can therefore study the problem of
cosmological structure formation in the non-expanding limit.
It is a natural starting point for a statistical mechanics
approach to this problem, and as we will see one which already
has most of the qualitative features of the cosmological problem.

\subsection{Relation to other $N\rightarrow \infty$ limits}

It is useful to comment on the relation of the (usual) thermodynamic limit 
we have discussed  to other limiting procedures used for the 
description of self-gravitating systems studied in the literature.  
More specifically we have in mind two such limits:
\begin{itemize}
\item the ``dilute'' thermodynamic limit, and
\item the Vlasov-Poisson limit. 
\end{itemize}

The first of these limits arises in the context of the well-known
treatment of gravity in the framework of the microcanonical ensemble,
which has been discussed by a series of authors, going back to Emden 
in the early twentieth century (see e.g. \cite{Padmanabhan:1989gm}
for a review, or the contribution of Chavanis in this
volume).  Such a treatment of a self-gravitating system of particles 
becomes possible if the particles are enclosed in a finite box 
and a cut-off is introduced to regularise the gravitational potential 
at the origin. Equilibrium states, i.e. states  maximising the 
entropy exist for sufficiently large energy, which may then survive
as meta-stable equilibria for the case that the cut-off goes to
zero. This treatment becomes formally valid in the limit
\be
\label{TDlimit-dilut}
N \rightarrow \infty, \, V \rightarrow \infty, \,\, {\rm at}
\,\, \frac{N}{V^{1/3}}={\rm constant} \,.
\ee
The mean mass and number density thus scale as
\be
\label{TDlimit-dilute-scaling}
\rho_0 \propto n_0=\frac{N}{V} \propto V^{-2/3}\,.
\ee
This scaling is clearly different to the limit we have
taken, at fixed mean mass and number density. This 
means that the results obtained from this microcanonical 
treatment are not expected to be relevant to the dynamics 
we will study. This can 
be seen by noting that the characteristic time scale 
for the dynamical evolution of the systems we study (and
which we will in fact use as our time unit below), is 
the {\it dynamical time}, defined as
\begin{equation}
\label{td}
\tau_{\rm {dyn}} \equiv \frac{1}{\sqrt{4\pi G \rho_0}}.
\end{equation}
This {\it physical} time scale\footnote{E.g. for a mass density of one
gram per cm$^3$, one finds $\tau_{\rm {dyn}} \approx 18.2$ minutes.} 
thus goes to infinity in the ``dilute'' thermodynamic limit, i.e. the
dynamics we describe never takes place. Alternatively we can say
that the effects described by this ``dilute'' limit are strictly
relevant to the systems we study only on times scales which are 
infinitely long compared to those which we consider. 
Indeed the physics described by this microcanonical treatment is 
understood to be relaxational, characterised by time scales which
diverge with $N$ when expressed in terms of the dynamical time-scales.
As we are studying already an infinite $N$ limit, this relaxational
time scale diverges, and any result from this treatment of a finite, closed,
system cannot describe the physics of our infinite system, of
which any finite subsystem is always open\footnote{This treatment may,
nevertheless, be relevant, qualitatively, to understanding aspects 
of the very long 
time evolution of the systems we study. Indeed, as we will 
describe, highly clustered regions can behave over much
longer timescales than $\tau_{\rm dyn}$ as roughly 
independent from the rest of the system, and so may manifest long-time
relaxational behaviors like those of isolated finite systems which are,
at least partially, described by this microcanonical treatment.}.

The Vlasov-Poisson limit, on the other hand, is a limit which has 
been studied for finite systems (see e.g. \cite{spohn}).
Its validity as an approximation to the full dynamics
of an $N$-body system has been rigorously demonstrated \cite{braun+hepp},
for gravity {\it with a regularisation} of the 
singularity in the potential, when the limit $N\rightarrow \infty$
is taken at fixed volume and a fixed time, with the
scaling of the particle mass $m \propto 1/N$. This latter
scaling makes the acceleration given by the sum on the
right hand side in Eq.~(\ref{Nbody-eom}) converge 
to the appropriate mean field value. In this case we therefore have
\be
\label{VPlimit-scaling}
n_0=\frac{N}{V} \propto N \,\,{\rm and}\,\,\rho_0=\frac{mN}{V}={\rm constant}\,.\ee
Again this is clearly a different limit to the one we have 
considered.  We see, however, that this limit leaves invariant
the characteristic time scale $\tau_{\rm dyn}$, which we noted
is that of the dynamics for the infinite systems we will describe
and analyse below. The reason for this is in fact that this time
scale emerges from a treatment of this dynamics which is derived 
by treating it in a Vlasov-Poisson approximation. In this case, however, this 
approximation now applies to the infinite system, i.e.,  
in which the limit $N \rightarrow \infty$ has already been
taken, but as we have described, at constant $n_0$ and
$\rho_0$. Although there is no rigorous demonstration in the
literature of the validity of such a limit for such a 
system, it is clear that it must correspond, if it indeed
exists, to the limit
\be
n_0 \rightarrow \infty\,\, {\rm at} \,\, \rho_0=mn_0={\rm constant}\,.
\ee
Indeed when the Vlasov limit is taken in a fixed (finite) volume
it may be written in this volume-independent way, which thus
generalizes without difficulty to the infinite volume case.
As we will discuss further below, in cosmology this limit
is of great importance, as the {\it aim} of the numerical 
simulations of self-gravitating Newtonian particle systems 
in this context is to reproduce this limit as well as possible.

Lastly we emphasize an important point concerning the usual 
thermodynamic limit for self-gravitating particles: we have shown that 
this limit exists for the {\it dynamical problem}. It is well known, and indeed
one of the starting points of any discussion of long-range interactions,
that in this limit the use of the usual instruments of equilibrium
thermodynamics is problematic. Indeed the gravitational potential
violates both the essential conditions necessary for their use
formulated by Ruelle\cite{ruelle} ---  the condition of ``temperedness'' due to
the slow decay of the potential (i.e. slower than the dimensionality 
of space) at large separations, and the condition of stability  
due to the behaviour at small separations. Our treatment of the
problem in the usual thermodynamic limit is completely consistent
with this. Indeed it turns that the dynamical problem we have
formulated is an intrinsically out of equilibrium one, the
evolution of the system remaining always explicitly time
dependent (in a calculable manner, determined as we will
see, by the analysis in the Vlasov-Poisson limit). Thus
any treatment of the problem as an equilibrium one, or
a quasi-equilibrium one (see e.g. the contribution of Saslaw 
in this volume) will at best be a useful, but non-rigorous, 
approximation.

\section{Evolution from Shuffled Lattice initial conditions: phenomenology}
\label{sec2} 

We describe in this section, in a phenomenological manner, the
evolution under their self-gravity of particles initially placed 
in a shuffled lattice (SL) configuration as observed in a set of
numerical simulations. This synthesizes the salient results of a
more detailed study reported in \cite{sl1}. Having given
the exact definition of a SL, and explained our choice of
this specific class of initial conditions, we will present
the results of simulations, focusing on the standard two point
characterisations of the observed evolving correlations, in terms of the 
reduced two point correlation function and its reciprocal space 
equivalent, the power spectrum\footnote{We will recall the definitions of 
these quantities and some of their essential properties at the 
appropriate points, but not the details of how they are estimated in 
the simulations. These may be found in \cite{sl1}.}. In the 
presentation of these
results we will highlight the qualitative similarity with
the known behaviour of analogous cosmological simulations,
and explain for uninitiated readers the essential
theoretical results --- just the original
analysis of Jeans of the evolution of density perturbations
in the fluid limit --- used to understand them.

\subsection{Shuffled Lattices}
\label{sec:charact}

We use the term SL to refer to the infinite point distribution
obtained by randomly perturbing a perfect cubic lattice: each 
particle on this lattice, of lattice spacing $\ell$, is moved 
randomly (``shuffled'') about its lattice site {\it
independently} of all the others.  A particle initially at the lattice
site $\ve R$ is then at $\ve x(\ve R) = \ve R + \ve u(\ve R)$, where
the random vectors $\ve u(\ve R)$ are specified by $p(\ve u)$, the 
PDF for the displacement of a single particle. 

An ensemble of SL configurations is thus characterised by the 
lattice spacing $\ell$ and the PDF $p(\ve u)$. 
We will use in the simulations below a simple top-hat form,
i.e., 
$p(\bu) = (2\Delta)^{-3}$ for $\bu \in [-\Delta,\Delta]^3$,
and $p(\bu)=0$ otherwise\footnote{We will explain below that
the exact form of the PDF is not of importance.}. 
Taking $\Delta\to 0\,$, at fixed $\ell$, one thus obtains a
perfect lattice, while taking $\Delta\to \infty$ at fixed $\ell$, 
gives an uncorrelated Poisson particle distribution\cite{book}.
Once the form of the PDF is specified the class of SL we 
consider is thus a two parameter family, characterized by
$\ell$ and $\Delta$. We refer to the latter as the \emph{shuffling parameter}.
Using the top-hat PDF it is simple to verify that it is 
simply the square root of the variance of the displacement
PDF, i.e., $\Delta^2 = \int d^3 u   {\ve u}^2 p (\ve u )$.
It is convenient to define also the dimensionless ratio 
$\delta \equiv \Delta/\ell$, which we will refer to as
the \emph{normalized shuffling parameter}. 

To characterize the correlation properties of the SL let us consider
its power spectrum. We recall that for a point (or continuous mass)
distribution in a cube of side $L$, with periodic boundary conditions,
this quantity is defined as\footnote{We adopt the
normalisation which is used canonically in cosmology,
for which the asymptotic behaviour at large
$\bk$, characteristic of any point process,
is $P( \ve k \rightarrow \infty )=\frac{1}{n_0}$.}
\begin{equation}
P(\ve k) = \frac{1}{L^3}\langle | \tilde\delta(\ve k) |^2 \rangle \,,
\label{eq:pktheo}
\end{equation}
where $\tilde\delta(\ve k)$ are the Fourier components of the 
density contrast  $\delta(\bx)\equiv (\rho(\bx) - \rho_0)/\rho_0$
(and $\rho(\bx)$ is the mass density), i.e., 
\begin{equation}
\tilde\delta(\ve k) =\int_{L^3} \delta(\ve x)
\exp(-i\ve k\cdot \ve x) \ \D[3] \ve x \,. 
\end{equation}
for $\ve k =(2\pi/L) \ve n$ where $\ve n$ is
a vector of integers. The infinite volume limit is
then obtained by sending $L \rightarrow \infty$, keeping
the average mass (or number) density fixed. We give 
the finite volume form for $P(\bk)$ here as we will
in fact treat in our simulations an infinite system
defined in this way.

It is straightforward (see Ref.~\cite{book}) to show that
the power spectrum for a generic SL is given by
\begin{equation}
 P(\ve k) = \frac{1-|\tilde{p}(\ve k)|^2}{n_0} + L^3 \sum_{\ve n \neq 0}
 |\tilde{p}(\ve k)|^2 \,  \delta_{\rm{K}}( \ve k , \ve n \frac{2 \pi}{\ell}) 
\label{eq:exactPS}
\end{equation}
where $\tilde p(\ve k)$ is the Fourier transform of the 
PDF for the displacements $p(\ve u)$ (i.e. its characteristic 
function),  and $\delta_{\rm{K}}$ is the three-dimensional 
Kronecker symbol. 
The second term in
Eq.~(\ref{eq:exactPS}) is non-zero only at the corresponding
discrete (non-zero) values of $\bk$. It is in fact simply the 
power spectrum of 
the unperturbed lattice modulated by $|\tilde{p}(\ve k)|^2$.
Thus only the first term in
Eq.~(\ref{eq:exactPS}) contributes around $\bk=0$. Taylor expanding
this term, assuming only that $p(\bu)$ has finite variance,
we obtain the leading small-$k$ behavior
\begin{equation}
P(\ve k) \approx \frac{k^2 \Delta^2}{3n_0} =\frac{1}{3} k^2 \delta^2
 \ell^5  \,,   
\label{eq:pk}
\end{equation}
where $k=|\bk|$. Note that this small-$k$ behavior of the 
power spectrum of the SL therefore does not depend on the 
details of the chosen PDF for the displacements,
but only on its (finite) variance. For this reason in fact
the results we will discuss for the chosen PDF do not
in fact depend in detail on this specific 
form\footnote{In Ref.~\cite{sl2} simulations like those discussed here,
but with a Gaussian PDF are analysed.}. Note that the power spectrum is 
a function only of $k$ at small $k$,
which means that the large scale fluctuations in the
system are statistically isotropic. The 
behavior $P(k) \propto k^2$ means that, as one 
would expect, an SL is at large scales a distribution with 
fluctuations which are suppressed compared to those in a Poisson 
distribution. Indeed the SL belongs for this reason 
to the class of ``superhomogeneous''\cite{glasslike} 
(or ``hyperuniform''\cite{to03}) point processes with this
property, characterized by the behavior 
$ P(\bk \rightarrow 0)=0$.

Let us now comment on why we choose this particular class
of SL initial conditions. The reason is that they are a very simple
(two parameter) class of initial conditions resembling those 
used in cosmological simulations. In this context initial conditions 
are prepared by applying {\it correlated} displacements to a lattice.
By doing so one can produce, to a good approximation, a desired
power spectrum at small wavenumbers\footnote{See \cite{discreteness1_mjbm}
for a detailed analysis of the algorithm used.}. The spectra of 
currently favored
models contain several parameters, roughly interpolating between
power law forms $P(k) \propto k^n$, with $n$ varying from $+1$
at the smallest $k$ to close to $-3$ at the largest $k$ included
in the initial conditions of numerical simulations. In particular
the SL allows us to mimic an important feature
of these simulations, related to the problem of discreteness 
we will discuss further below: with the two parameters characterising
the class of initial conditions, {\it we can modify independently the
particle density and the amplitude of the power spectrum at small
wavenumbers.} This can be seen from Eq.~(\ref{eq:pk}), as it
is clear that to keep the latter fixed it suffices to impose,
when varying $\ell$, the condition
\begin{equation}
\delta^2 \ell^5 ={\rm constant}\,.
\label{condition}
\end{equation}

\subsection{Numerical Simulations}

In numerical simulations we cannot treat of course a truly 
infinite SL. Instead we treat the infinite
system defined by such a lattice, in a finite cubic box of 
size $L=N^{1/3} \ell$, plus periodic copies. Our results will 
then be representative 
of the thermodynamic limit only insofar as they do not depend 
on the size of the box. We will see below that the nature 
of the dynamics is such that this can be true only 
for a finite time, i.e., the evolution of this 
infinite, but periodic, system will represent that in the 
thermodynamic limit for a finite time. To follow the evolution 
for a longer time requires an ever larger box. Any effects which
depend on the box size are {\it finite size} effects which are not
related to the real behavior of the well defined 
physical limit we are studying\footnote {A good analogy is in
the simulation of the out of equilibrium dynamics of glassy
systems, e.g., the ordering dynamics of a quenched ferromagnet.}.

In Table~\ref{tab:sl_summary} are given the details 
of the SL initial conditions of the N body simulations
reported in \cite{sl1}, and which we will discuss here. 
\begin{table}
\begin{tabular}{cccccccc}
Name & $N^{1/3}$ & $L$ & $\ell$ & $\Delta$ & $\delta$ & $m/m_{64}$ \\ 
\hline 
SL64 & 64 & 1& 0.015625& 0.015625 & 1       & 1     \\ 
SL32 & 32 & 1&0.03125  & 0.0553   & 0.177   & 8     \\ 
SL24 & 24 & 1&0.041667 & 0.00359  & 0.0861  & 18.96 \\ 
SL16 & 16 & 1&0.0625   & 0.00195  & 0.03125 & 64    \\ 
\hline 
SL128& 128&2 &0.015625 & 0.015625 & 1       & 1     \\ 
\end{tabular}
\caption{Details of the SL used as initial conditions in the
  simulations of \cite{sl1}. $N$
  is the number of particles, $L$ is the box size, $\ell$ the lattice
  constant and $\Delta$ ($\delta$) the (normalized) shuffling parameter.
  \label{tab:sl_summary} }
\end{table}

The numerical evolution has been performed using 
the publicly available code \textsc{Gadget} \cite{gadget,gadget_paper}. 
This is based on a tree algorithm for the calculation of the force, and
allows one to perform simulations in an infinite periodic 
distribution using the Ewald summation method.  
The potential used is {\it exactly} equal to the Newtonian potential 
for separations greater than a softening length $\varepsilon$, and
regularized below this scale\footnote{We do not give here the 
rather complicated functional form of this smoothing, which can be found 
in \cite{gadget,gadget_paper}. The important point here is that the 
smoothed force goes to zero continuously as particles approach one another.}.
More details of tests on the accuracy of the 
simulations (energy conservation, robustness to change of
integration parameters etc.) can be found in \cite{sl1}.

\begin{figure}
\includegraphics[width=0.5\textwidth]{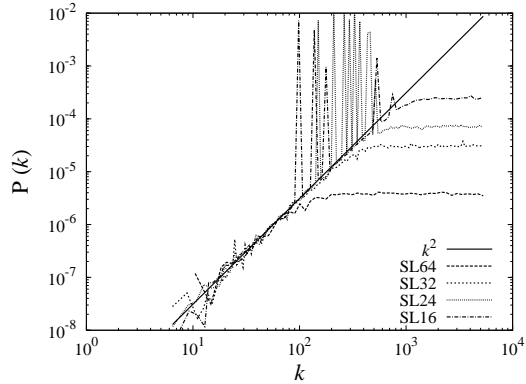}
\caption{The power spectra (averaged in spherical shells) of
the SL configurations
specified above in Tab.~\ref{tab:sl_summary} as a function of the 
modulus of $\ve k$. The
solid line is the theoretical ($\propto k^2$) behavior for small $k$
given by Eq.~\eqref{eq:pk}. At large $k$, the four power spectra
are equal to $1/n_0$, with the corresponding value of $n_0$. The peaks arise from
the second term in Eq.~\eqref{eq:exactPS}. From Ref.~\cite{sl1}. 
\label{fig:pk_t0}}
\end{figure}

Let us explain the rationale for the choice of the parameters 
in Table~\ref{tab:sl_summary}.  Firstly, we have chosen 
our (arbitrary) units of length, mass and time as follows. 
Our unit of length is given by the box side of the SL64
simulation and our unit of mass by the particle mass in this same
simulation. The particle masses are then chosen so that 
the mass density $\rho_0$ is constant, which means that the
dynamical time defined above in Eq.~(\ref{td}) is the same 
in all simulations. This is convenient because this is, as
we have mentioned above, an appropriate characteristic
time-scale for the evolution which can be derived from the
fluid limit (see below).
We have chosen a smoothing parameter $\varepsilon = 0.00175$
in all simulations. It is therefore in all cases significantly 
smaller --- at least a factor of ten ---  than the lattice
spacing (which is close in all cases to the initial average 
distance between nearest neighbors). The box size is 
the same in all but one simulation. This latter 
simulation (SL128), which is the biggest one, is used to 
test the accuracy with which our results are
representative of the thermodynamic limit. Thus it is 
chosen to have the same parameters to SL64, differing only in its volume 
(which increases by a factor of 8). Finally the values
of $\delta$ have been chosen for each $\ell$ so that
the condition in Eq.~(\ref{condition}) is fulfilled.
The measured power spectra of realizations of SL with
the parameters given in Table~\ref{tab:sl_summary} are shown in
Fig.~\ref{fig:pk_t0}. We see, up to statistical fluctuations, that the
spectra are indeed of the same amplitude at small $k$. Note that the
curve corresponding to each initial condition is most easily
identified by the fact that the asymptotic (large $\bk$)
level of the power spectrum is inversely proportional to the mean
particle density.

Lastly, the particles are assigned zero velocity in the initial
conditions (at $t=0$), and they are run in each case until a time
when the finite size of the box manifestly becomes important
(see below).

Let us comment further on the use of the smoothing parameter
$\varepsilon$. The goal here is to study the evolution of a
distribution of self-gravitating {\it point} particles, and 
the reason for the introduction of $\varepsilon$ is that it 
allows a much more rapid numerical integration. It does this 
because it modifies
the trajectories in which particles have close encounters,
which in lead to very large accelerations
(and thus the necessity for very small time steps).
The use of the smoothing is only justified therefore if the
associated physical effects are not important for the 
macroscopic properties we will study. A priori we do not
know whether this is the case, and we simply test numerically
(see \cite{sl1}) for the robustness of our results to changes, towards 
significantly smaller values, of the smoothing parameters.
We interpret the observed $\varepsilon$-independence of our
results as meaning that the artificial
effects it introduces in the dynamics may be relevant 
to the quantities we measure only on time scales longer than 
those we can study in our simulations.

We mention in this context the following point, to which we will
return further below. In cosmological simulations of particle 
systems the smoothing is understood as a physical parameter,
introduced with the intention of making the particle system 
behave more like the desired Vlasov-Poisson limit 
(in which two body encounters are neglected).
In practice, however, values of the ratio $\epsilon/\ell$ even
smaller than those adapted here are typically used. Thus the 
parameter choices we have made here would be considered as 
perfectly reasonable if our goal were, as in cosmology, to reproduce 
the Vlasov-Poisson limit (and, as far as possible, nothing else). 

\subsection{Results}

\begin{figure}
\includegraphics[width=0.75\textwidth]{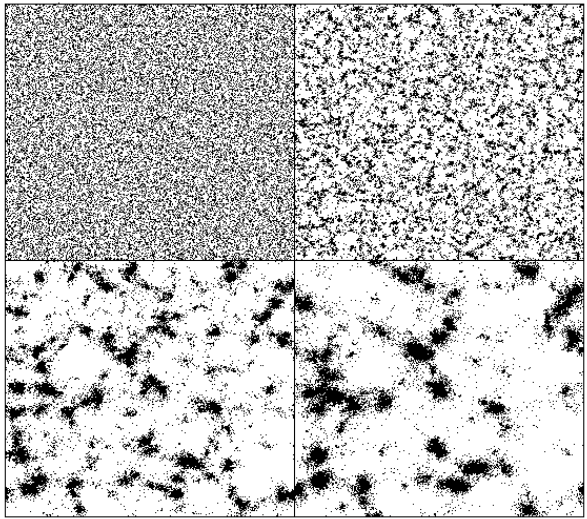}
\caption{Snapshots of SL64, at $t=0$, and the evolved
configurations obtained at subsequent times, t=$3,6,8$.
These are projections on the $z=0$ plane. 
\label{fig:snapshots}}
\end{figure}

In Fig.~\ref{fig:snapshots} are shown four 
snapshots\footnote{The results taken from Ref.~\cite{sl1}
are given in
units which are not exactly those of dynamical time, but rather in which
$\tau_{\rm {dyn}}=1.092$. This corresponds to time units 
of $1000$ seconds if the mass density were $1$g.cm$^{-3}$.}
of the simulation SL64. We see visually that non-linear structures
(i.e. regions of strong clustering) appear to develop first at
small scales, and then propagate progressively to larger scales.
Eventually the size of the structures become comparable to the
box-size. From this time on the evolution of the system will
no longer be representative of the thermodynamic limit we are
studying. Up to close to this time, however, it is indeed the
case that all the properties we will study below show negligible
dependence on the box size (see Refs.~\cite{sl1,sl2} for more detail).

\subsubsection{Evolution of correlations in real space}
 
We consider now the evolution of clustering in real space, as
characterized by the reduced correlation function $\xi(r)$. 
We recall that this is simply defined as 
\begin{equation} 
\xi(\ve r)= \langle \delta(\ve x+\ve r)\delta(\ve x)\rangle
 \;,
\end{equation}
where $\langle ... \rangle$ is an ensemble average, i.e., an average
over all possible realizations of the system (and we have assumed
statistical homogeneity). It is useful to note 
that this can be written, for $r\neq0$, and averaging over spherical
shells, 
\begin{equation}
\xi(r) = \frac{\langle n(r)\rangle_p }{n_0}-1 
\;,
\label{eq:xi_rdf}
\end{equation}  
where $\langle n(r)\rangle_p$ is the \emph{conditional average density},
i.e., the  (ensemble) average density of points in an infinitesimal
shell at distance $r$ from a point of the distribution. 
Thus $\xi(r)$ measures clustering by telling us how the density 
at a distance from a point is affected, in an average sense, by 
the fact that this point is occupied.  In distributions which 
are statistically homogeneous the power spectrum $P(\bk)$
and $\tilde{\xi} (\ve r)$ are a Fourier conjugate 
pair (see e.g. Ref.~\cite{book}).

The correlation functions we will discuss below
will invariably be monotonically decreasing
functions of $r$. It is then natural to define the 
scale $\lambda$ by  
\begin{equation}
\xi(\lambda) = 1  \,.
\label{eq:homoscale}
\end{equation}
This scale then separates the regime of weak correlations
(i.e. $\xi(r) \ll  1$) from the regime of strong 
correlations (i.e. $\xi(r) \gg  1$). In the context of gravity 
these correspond, approximately, to what are referred
to as the {\it linear} and {\it non-linear} regimes,
as a linearized treatment of the evolution of
density fluctuations (see below) is expected to be valid 
in the former case.  Eq.~(\ref{eq:homoscale}) can also 
clearly be considered as a definition of 
the \emph{homogeneity scale} of the system. Physically
it gives then the {\it typical size of strongly clustered
regions}. 

\begin{figure}
\includegraphics[width=0.5\textwidth]{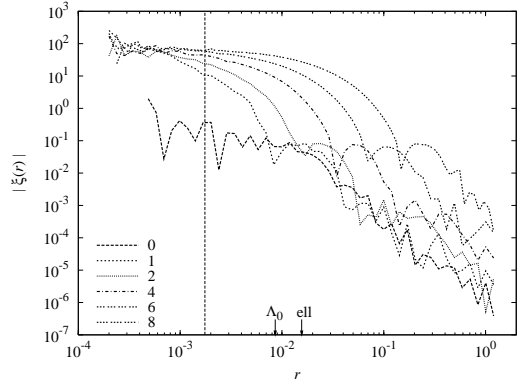}
\caption{Behavior of the absolute
value of the correlation function $|\xi(r)|$ in SL128 at times
$t=0,1,2,4,6,8$. The vertical dotted line indicates $\varepsilon$. 
From \cite{sl1}.}
\label{fig:xi}
\end{figure}

In Fig.~\ref{fig:xi} is shown the evolution of the absolute value 
$|\xi(r)|$ in a log-log plot, for the SL128 simulation.
These results translate quantitatively the visual impression
gained above. More specifically we observe that:
\begin{itemize}
\item Starting from $\xi(r) \ll 1$ everywhere, non-linear correlations
(i.e. $\xi(r)\gg 1$ ) develop first at scales smaller than the
initial inter-particle distance.

\item After two dynamical times the clustering
develops little at scales below $\varepsilon$. The clustering around
and below this scale is characterized by an approximate ``plateau''.
This corresponds to the resolution limit imposed by the chosen
smoothing.

\item At scales larger than $\varepsilon$ the correlations
grow continuously in time at all scales, with the scale of
non-linearity [which can be defined, as discussed above, by
$\xi(\lambda)=1$] moving to larger scales.

\end{itemize}

From Fig.\ref{fig:xi} it appears that, once significant non-linear
correlations are formed, the evolution of the correlation function
$\xi(r)$ can be described, approximately, by a simple ``translation''
in time. This suggests that $\xi(r,t)$ may satisfy in this regime a
spatio-temporal scaling relation:
\begin{equation}
\xi(r,t) \approx \Xi\left( r/R_s(t) \right)\,, 
\label{eq:rescaling}
\end{equation}
where $R_s(t)$ is a time dependent length scale which we discuss in
what follows.
\begin{figure}
\includegraphics[width=0.5\textwidth]{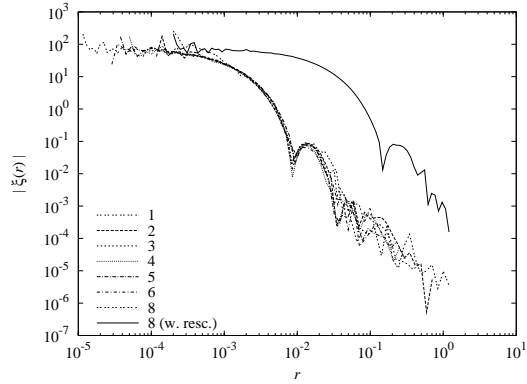}
\caption{Collapse plot of $\xi(r,t)$: for each time $t>1$ we 
have rescaled the $x$-axis by a time-dependent factor to collapse all
the curves (dashed ones) to that at time $t=1$. We have added for
comparison $\xi(r,t=8)$ without rescaling (``w. resc.'', continuous
line). From \cite{sl1}.}
\label{fig:collapse}
\end{figure}
In order to see how well such an ansatz describes the evolution, 
we show in
Fig.~\ref{fig:collapse} an appropriate ``collapse plot'': $\xi(r,t)$
at different times is represented with a rescaling of the $x$-axis by
a (time-dependent) factor chosen to superimpose it as closely as
possible over itself at $t=1$, which is the time from which the
``translation'' appears to first become a good approximation.  We can
conclude clearly from Fig.~\ref{fig:collapse} that the relation
Eq.~(\ref{eq:rescaling}) indeed describes very well the evolution, down to
separations of order $\varepsilon$, and up to scales at which
the noise dominates the estimator (see \cite{sl1} for further details).
\begin{figure}
\includegraphics[width=0.7\textwidth]{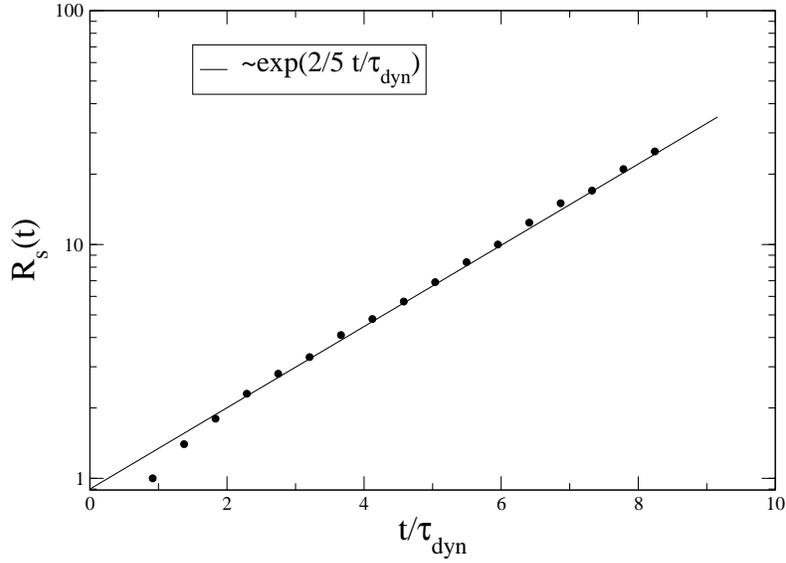}
\caption{Evolution of the function $R_s(t)$ in SL128
(points) compared with its prediction from linearized
fluid treatment, as explained in the next section.
From \cite{sl1}.}
\label{fig:Rst}
\end{figure}
In Fig.~\ref{fig:Rst} is shown the evolution of the rescaling factor
$R_s(t)$ found in constructing Fig.~\ref{fig:collapse}, as a function
of time [with the choice $R_s(1)=1$]. Also shown is
a theoretical curve which we will explain in the next subsection.
Note that since we have defined the homogeneity scale
$\lambda$ by $\xi(\lambda)=1$ it is clear that, once the
spatio-temporal scaling relation is valid, we 
have $\lambda(t) \propto R_s(t)$. A fit to a simple functional 
form for $\Xi(r)$, a shallow power law followed
by an exponential cut-off, may be found in Ref.~\cite{sl1}.


\subsubsection{Evolution of the Power Spectrum} 
\label{ps-evolution}

To see how the observed temporal dependence of the
correlation function is explained theoretically, in analogy 
to how this is done in cosmological simulations, we outline very
briefly the essential result of the Jeans-type analysis
of the evolution of perturbations in a self-gravitating,
and pressureless, fluid. For comparison
of its predictions with the numerical results for the
discrete N-body system, it is simplest then to turn to
the reciprocal space description of the clustering
\footnote{The real and reciprocal space description
are, of course, fundamentally equivalent. The convenience
of one space over another is simply in how effects 
due to their estimation in finite samples enter.}.
 
The equations describing the evolution of a 
non-relativistic pressureless self-gravitating fluid are the 
following\footnote{As mentioned above a rigorous derivation
of these equations in the thermodynamic limit as we have defined
it can be found in Ref.~\cite{kiessling}.}
\bea
\partial_t \rho + \nabla_{\ve x}\cdot (\rho \ve v) = 0, \qquad
\partial_t  \ve v  + (\ve v \cdot \nabla_{\ve x})  \ve v 
=\ve g\ \nonumber \\
\nabla_{\ve x} \cdot \ve g = -4\pi G ( \rho -\rho_0 ), 
\qquad \nabla_{\ve x} \times \ve g = 0 \ ,
\nonumber
\eea
where $\rho(\ve x,t)$ is the mass density, $\ve v(\ve x,t)$ the
velocity field and $\ve g(\ve x,t)$ the gravitational field.
These equations can also be obtained by an appropriate truncation of 
Vlasov-Poisson equations. Linearizing in the density 
perturbations, and the velocity, one obtains 
\begin{equation}
  \ddot{\delta}(\ve x,t) = 4\pi G \rho_0 \delta(\ve x,t) \,\, {\rm and}\,\, 
  \ddot{\tilde\delta}(\ve k,t) = 4\pi G \rho_0 \tilde\delta(\ve k,t) \ .
\label{eq:ddotdelta}
\end{equation}
For the case we are analysing, of zero initial velocities, therefore
\begin{equation}
  \delta(\bk,t) = \delta(\bk,0) \cosh\left( \sqrt{4\pi G \rho_0}\
  t\right) \;. \label{eq:densitycontrastlinear} 
\end{equation}
and thus for the power spectrum
\begin{equation}
P(\ve k,t) = P(\ve k,0) \cosh^2\left(t/\tau_{\rm {dyn}} \right) \ .
\label{eq:pk_evolution_linear}
\end{equation}

\begin{figure}
\includegraphics[width=0.5\textwidth]{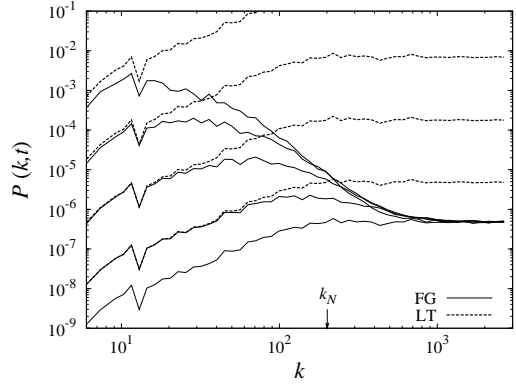}
\caption{Evolution of the power spectrum 
in SL128 (solid lines --- label FG): the
  curves are for time equal to 0,2,4,6,8 (from bottom to up).  The
  dashed lines labeled with LT show the predictions of fluid linear
  theory, i.e., Eq.~\eqref{eq:pk_evolution_linear} with $P(\bk,0)$
  measured in the simulation at $t=0$ for the same time steps. The
  arrow labeled ``$k_N$'' shows the value of the corresponding Nyquist
  frequency $k_N = \pi/\ell$. From \cite{sl1}.
\label{fig:pk_SL128} }
\end{figure}

The evolution of the power spectrum as estimated in SL128 is 
shown in Fig.~\ref{fig:pk_SL128}. Along with the numerical 
results is shown the prediction Eq.~(\ref{eq:pk_evolution_linear}).
We observe that:
\begin{itemize}
\item 
The linear theory prediction describes the evolution very accurately
in a range $k < k^{*}(t)$, where $k^{*}(t)$ is a wave-number which
decreases as a function of time. This is precisely the qualitative
behavior one would anticipate (and also observed in cosmological
simulations): linear theory is expected to hold approximately
only above a scale in real space, and therefore up to some corresponding 
wavenumber in reciprocal scale, at which the averaged density 
fluctuations are sufficiently small so that the linear approximation 
may be made\footnote{A more precise study of the
validity of the linearized approximation is given in \cite{sl1}.}.
This scale in real space, as we have seen, clearly 
increases with time,  and thus in reciprocal space decreases with 
time. We note that at $t=6$ only the very smallest $k$-modes 
in the box are still in this {\it linear regime}, while at $t=8$ 
this is no longer true (and therefore finite size effects are expected
to begin to play an important role at this time).

\item
At very large wave-numbers, above $k_N$, the power spectrum remains 
equal to its
initial value $1/n_0$. This is simply a reflection of the necessary
presence of shot noise fluctuations at small scales due to the
particle nature of the distribution. 

\item 
In the intermediate range of $k$ the
evolution is quite different, and {\it slower}, than that given by
linear theory. This is the regime of {\it non-linear clustering}
as it manifests itself in reciprocal space.
\end{itemize} 



\subsubsection{Spatio-temporal scaling and ``self-similarity''}

We now return to real space and explain a little further how
the spatio-temporal scaling relation Eq.~(\ref{eq:rescaling})
for the correlation function we have observed may be linked
to the linearized fluid theory, and then what it tells us
about the nature of the clustering in the system.

In the context of cosmological N body simulations this kind of
behavior, {\it when} $R_s(t)$ {\it is itself a power law (in time)},
is referred to as {\it self-similarity}. 
Such behavior is expected in
an evolving self-gravitating fluid (see
e.g. \cite{peebles,efstathiou_88,bertschinger}) because of the
scale-free nature of gravity, {\it if} the expanding universe model
{\it and} the initial conditions {\it contain no characteristic
scales}, i.e., if the input model has a simple power law
$P(\bk) \propto k^n$. On theoretical grounds there are different expectations
(\cite{bertschinger,bertschinger2,efstathiou_88}) about the range of
exponents $n$ of the power spectrum which should give self-similar
behavior, because of how effects coming notably from 
particle discreteness, the finite box size and force smoothing
break the scale-free idealisation. It is widely agreed that
it applies well for $-1<n<-1$ in the literature, but there
are differing conclusions about the observed degree of 
self-similarity outside this range.

The simple arguments used in this context to predict the temporal
scaling of the correlation functions can be generalized easily
to the case of a non-expanding universe (which, just like the
relevant cosmological models, has no characteristic scale).
To do so one simply {\it assumes} that such a spatio-temporal 
scaling relation holds exactly, {\it i.e., at all scales}, 
from, say, a time $t_{s} > 0$. For $t > t_{s}$ we have then 
\bea
P(\ve k,t) &=&  \int_{L^3} \exp(-i\ve k\cdot \ve r) \,{\xi}(\ve
r,t) \, \D[3] \ve r \\
&=&  R_s^3(t) \int_{L^3} \exp(-i R_s(t) \ve k\cdot \ve x ) \,
\Xi(|\ve x|) \
\D[3] \ve x   \\
&=&  R_s^3(t) P(R_s(t)\ve k,t_s)  \ .
\label{eq:Rsdetermine}
\eea
where we have chosen $R_s(t_s)=1$. Assuming now that 
the power spectrum at small $k$ is amplified as given by linear theory,
i.e., as in Eq.~(\ref{eq:pk_evolution_linear}), one infers
for any power spectrum $P(k) \sim k^n$ that
\footnote{Linear theory is expected\cite{peebles} in fact to describe the
evolution of the power spectrum at small $k$ only for $n<4$.
This is because any reorganisation of smaller scales itself
produces a term $\propto k^4$, so that evolution on small
scales is then expected to dominate the evolution of
large scale modify that at small $k$.} 
\begin{equation} 
R_s(t) = 
\left( \frac{\cosh \frac{t}{\tau{\rm{dyn}}}}{\cosh \frac{t_s}{\tau{\rm{dyn}}}} \right)
^\frac{2}{3+n} 
\rightarrow
\exp\left[ \frac{2(t-t_s)}{(3+n)\tau{\rm{dyn}}} \right]
\,\, {\rm for} \,\, {t\gg t_s}\,.
\label{eq:predRst}
\end{equation}
In the asymptotic behavior the relative rescaling in space 
for any two times becomes a function only 
of the {\it difference} in time between them so that we can write
\begin{equation} 
\xi (r, t + \Delta t) = \xi \left(\frac{r}{R_s(\Delta t)}, t \right)
\; ; \quad R_s(\Delta t) = e^{ \frac{2 \Delta t} {(3+n) \tau{\rm{dyn}}} } \,.
\label{eq:self-sim-xi}
\end{equation}

Let us now return to Fig.\ref{fig:Rst}. In addition to the measured
values of $R_s(t)$ is plotted the best-fit (corresponding to the time
$t_s \sim 2.5$) given by Eq.~(\ref{eq:self-sim-xi}), with $n=2$ for the SL.
The agreement is clearly very good. The system is thus indeed
very well described by the self-similar scaling predicted
by linearized fluid theory after an initial transient time.

All these behaviours of the two point correlations are qualitatively
just like those observed in cosmological simulations in 
an expanding universe. More general than the ``self-similar''
properties (which apply only to power law initial conditions),
the clustering can be described as ``hierarchical'' , a feature
typical of all currently favoured cosmological (``cold dark matter'' 
-type) models: structures develop at a scale which increases in
time, at a rate which can be determined from linear theory. This
is given the following physical interpretation: clustering may be 
understood essentially as produced by the collapse of small initial 
overdensities which evolve as prescribed by linear theory,
independently of pre-existing structures at smaller scales,
until they ``go non-linear''. Thus the process is essentially
characterised by a flow of power from large scales to small
scales\footnote{The extrapolation of this kind of picture has been
used in constructing various quite successful phenomenological
models which has been used to describe the non-linear clustering 
observed in numerical simulations, notably 
the so-called ``halo-model'' (see e.g. Ref.~\cite{halo}) to 
which we will return below, as well as prescriptions like 
that of Peacock and Dodds \cite{peacock} for calculating the
final power spectrum.}. 

One of the crucial points about this description of clustering
is that it tends to support the hypothesis that cosmological 
simulations may be understood almost entirely within a 
fluid description, i.e., more broadly within the framework 
of Vlasov-Poisson equations. We will discuss this point further
below, after the next section, which shows that we can in fact
understand significant aspects of the clustering we have described 
without assuming this framework.


\section{Evolution from Shuffled Lattice IC: theory}

We present in this section various analytical and
numerical approaches to understanding more fully the dynamical 
evolution from SL initial conditions, beyond the linearized fluid
theory which we have considered in the previous section.

\subsection{Phase 1: ``Melting'' of the lattice}
\label{PLT Evolution of a perturbed lattice}

The evolution of a self-gravitating perturbed 
lattice --- whether these perturbations
are correlated or not --- may be described using a perturbative 
treatment completely analogous to that used in condensed matter
physics to describe classical phonons (see e.g. Ref.~\cite{pines}).
Substituting ${\bf r}_i(t)={\bf R} + {\bf u}({\bf R},t)$ directly
in Eq.~(\ref{Nbody-eom}), where ${\bf R}$ is the lattice vector 
of the $i$th particle and ${\bf u}({\bf R},t)$ its displacement,
one obtains, expanding each term in the force to linear order
in the displacements, 
\be
{\bf {\ddot u}}({\bf R},t) 
= - \sum_{{\bf R}'} 
{\cal D} ({\bf R}- {\bf R}') {\bf u}({\bf R}',t)\,. 
\label{linearised-eom}
\ee
The matrix ${\cal D}$ is the {\it dynamical matrix}.
For gravity it may be written 
\be
{\cal D}_{\mu \nu} ({\bf R} \neq {\bf 0})=
Gm\left(\frac{\delta_{\mu \nu}}{R^3}
-3\frac{R_\mu R_\nu}{R^5}\right), \,\,
{\cal D}_{\mu \nu} ({\bf 0})= 
-\sum_{{\bf R} \neq {\bf 0}} {\cal D}_{\mu \nu} ({\bf R})
\nonumber
\ee
where $\delta_{\mu \nu}$ is the Kronecker delta.
Note that a sum over the copies, associated with
the periodic boundary conditions, is implicit in
these expressions.

The Bloch theorem  for lattices tells us that ${\cal D}$
is diagonalized by plane waves in reciprocal space.
We define the discrete Fourier transform and its inverse by 
\bea
\label{def-discreteFT-tok}
{\bf {\tilde u}}({\bf k},t)&= \sum_{{\bf R}} e^{-i {\bf k}\cdot{\bf R}}
{\bf u}({\bf R},t) \\
\label{def-discreteFT-tor}
{\bf u}({\bf R},t)&= \frac{1}{N} \sum_{{\bf k}} e^{i {\bf k}\cdot{\bf R}}
 {\bf {\tilde u}}({\bf k},t)\,,
\eea
where the sum in Eq.~\eqref{def-discreteFT-tor} is over the first
Brillouin zone of the lattice, i.e., for a simple cubic 
lattice $\bk = \bn (2\pi/L)$, where $\bn$
is a vector of integers of which each component $n_i$ ($i=1,2,3$) 
takes all integer values in the range $-N^{1/3}/2 < n_i \leq N^{1/3}/2$. 
Using these definitions in Eq.~(\ref{linearised-eom}) we obtain 
\be 
{\bf \ddot{{\tilde u}}} ({\bf k},t)  = -{\tilde{\cal {D}}} ({\bf k}) 
{{\bf {\tilde u}}}({\bf k},t) 
\ee 
where ${\tilde{\cal{D}}} ({\bf k})$, the Fourier 
transform (FT) of ${\cal D} ({\bf R})$, is a
symmetric $3 \times 3$ matrix for each ${\bf k}$.

Diagonalising ${\tilde{\cal{D}}} ({\bf k})$ one can determine, 
for each ${\bf k}$,  three orthonormal eigenvectors 
${\bf e}_n ({\bf k})$ and their eigenvalues 
$\omega_n^2({\bf k})$ ($n=1,2,3$), which obey the
Kohn sum rule
\be
\sum_n \omega_n^2({\bf k}) = -4 \pi G \rho_0 \,.
\ee

Given the initial displacements and velocities 
at a time $t=0$, the dynamical evolution of the
particle trajectories is then given as  
\be
\label{eigen_evol}
\bu(\bR,t)=
\frac{1}{N}\sum_{\bk}\left[{\cal{P}}(\bk,t)\tbu (\bk,0)+
{\cal{Q}}(\bk,t) {\dot{\tbu}} (\bk,0)\right]e^{i\bk\cdot\bR}
\ee
where the matrix elements of the ``evolution operator'' 
$\cal{P}$ and $\cal{Q}$ are
\bea
\label{evol_operators}
{\cal{P}}_{\mu\nu}(\bk,t)=&\sum_{n=1}^3 U_n(\bk,t)(\mathbf{e}_n(\bk))_\mu(\mathbf{e}_n(\bk))_\nu\\
{\cal{Q}}_{\mu\nu}(\bk,t)=&\sum_{n=1}^3 V_n(\bk,t)(\mathbf{e}_n(\bk))_\mu(\mathbf{e}_n(\bk))_\nu\,.
\eea
and\footnote{When this treatment is 
generalized (see Ref.~\cite{marcos_06}) for 
cosmological simulations, only these mode functions change.}
\be
U_n({\bf k},t)= \cosh \left(\sqrt{\omega_n^2 (\bk)}t\right) \,\,,\,,
V_n({\bf k},t)= 
[\sinh \left(\sqrt{\omega_n^2 (\bk)}t\right)]/\sqrt{\omega_n^2 (\bk)}
\ee
and the convention for the square root is: $\sqrt{\omega^2}=|\omega|$ 
if $\omega^2 > 0$, and $\sqrt{\omega^2}=i|\omega|$ if $\omega^2 < 0$.
Thus, depending on the sign of $\omega^2_n(\bk)$ and the initial
conditions, each mode is either growing/decaying or 
oscillatory in nature.

To fully solve for the evolution in this approximation requires 
thus only the determination of the eigenvectors and eigenvalues of the 
N 3$\times$3 matrices ${\tilde{\cal{D}}} ({\bf k})$. This 
is straightforward to do and computationally inexpensive.
\begin{figure}
\resizebox{8cm}{!}{\includegraphics*{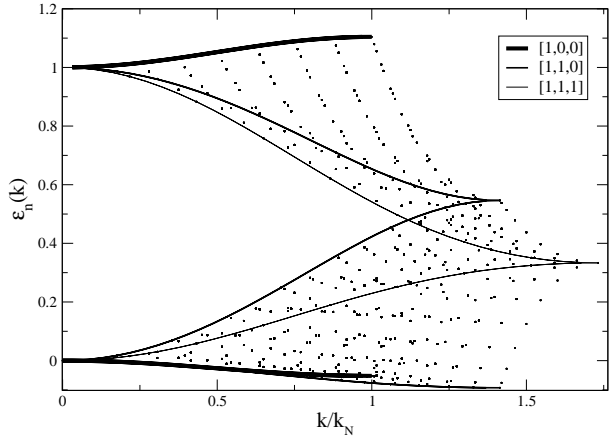}}
\caption{Normalized eigenvalues (see text) for the eigenvectors 
of the dynamical matrix of a $16^3$ simple cubic lattice.
From Ref.~\cite{marcos_06}.
\label{fig1}}
\end{figure}
In Fig.~\ref{fig1} are shown the spectrum of the normalized
eigenvalues 
\be \epsilon_n ({\bf k})\equiv -\frac{\omega_n^2({\bf k})}{4\pi G
\rho_0}\,,
\label{def-epsilon}
\ee
for a $16^3$ simple lattice. Full detail on these
calculations may be found in Ref.~\cite{marcos_06}.
The eigenvalues are plotted as a function of the modulus $k$ of
$\bk$, in units of the Nyquist frequency $k_N=\frac{\pi}{\ell}$. 
The fact that the eigenvalues at a
given $k$ do not have the same value is a manifestation of the 
anisotropy of the lattice. Shown in
the plot are also lines linking eigenvectors oriented in some chosen
directions. This allows one to see the branch structure of the
spectrum, which is familiar in the context of analogous calculations
in condensed matter physics (see e.g. \cite{pines}).  Increasing 
particle number only changes the density of reciprocal lattice vectors 
in these plots (since $\bk/k_N =
\bn/N^{1/3}$), just filling in more densely the plot of the eigenvalues
in Fig.~\ref{fig1} but leaving its form essentially unchanged.

Formally the expansion used of the force in this treatment
--- which we call ``particle  linear theory'' or PLT --- is 
valid until the relative separation of any two particles
becomes equal to their initial separation on the lattice,
but numerical study is required to determine the usefulness
in practice of the linearized approximation. In \cite{marcos_06} 
we have investigated this question in detail, comparing the 
results of the evolution under full gravity (in numerical simulations), 
with that obtained using the formulae just given. It turns out 
that it gives an excellent description of the 
evolution (for the typical quantities describing clustering) until 
a time when the average relative displacement approaches the 
lattice spacing, i.e., until a significant number of particles
start to approach close to one another (and the lattice ``melts'').

\subsection{Phase 2: Nearest neighbour domination}

The PLT approximation for the force breaks down when 
particles approach closely nearby particles, simply because
the force exerted becomes dominated by these particles. 
It is thus natural to consider whether this 
perturbative phase may be followed by one in which these
interactions between nearest neighbors (NN) may be 
relevant in understanding the clustering which emerges. 
It turns out that in fact a very good approximation to the evolution
of the SL is given by assuming an abrupt switch from
a phase in which the forces on particles are given by
the PLT approximation, to a phase in each particle feels
only the force due its NN. Further this approximation
works until a very significant amount of non-linear
correlation has developed, essentially covering most
of the transient period prior to the asymptotic self-similar
scaling we have discussed above.

In Refs.~\cite{sl1,sl2} indirect evidence was presented for
the validity of such a model, using the fact the following
relation was observed to hold in numerical simulations, 
in the relevant range of time scales:
\begin{equation} 
\label{eq:omega1} 
\omega(r) \, \D r = \left( 1 - \int_0^r\omega(s)\, \D s \right) \cdot
 4\pi r^2  \langle n(r)\rangle_p  \D r  \ , 
\end{equation}
Here $\omega(r)$ is the NN PDF, i.e., $\omega(r)dr$ is the probability
that a particle has its NN in the radial shell between $[r, r+dr]$.
The relation is exact\cite{book} if all but two point correlations may be
neglected, i.e., if the correlations may be fully accounted for
only due to two body correlations.

\begin{figure}[tpd] 
\scalebox{0.7}{\includegraphics*{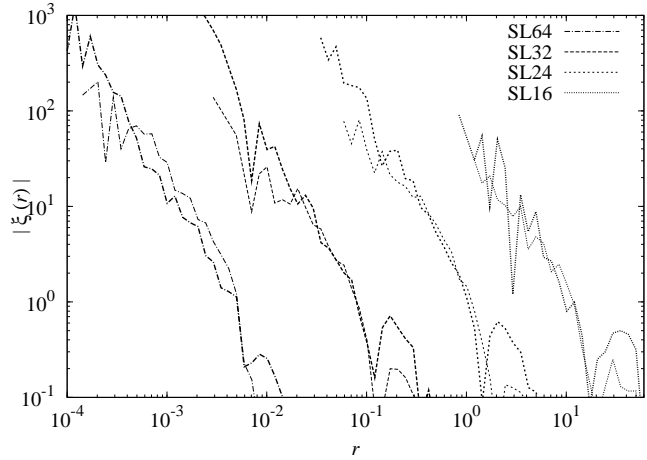}} 
\caption{
The two-point correlation function at the times
$t_{\rm max}=1,2.5,3.5,4.5$ for the different SL initial 
conditions as indicated,  in both the full gravity simulations (thick lines) 
and the simulations of the two phase model described in the text
(thin lines). 
For clarity the $x$ axis has been rescaled for each initial
condition (as otherwise the curves are, to a very good
approximation, all superimposed). From Ref.~\cite{sl3}. 
\label{fig:linear_NN}}
\end{figure}

In a recent paper \cite{sl3} we have shown directly the validity
of this two phase approximation to the evolution of an SL
at early times, by comparing the full numerical evolution
with gravity to a direct numerical integration of this model. Results are 
shown in Fig.~\ref{fig:linear_NN}. The only free parameter here is
the time at which we switch from integration with PLT to
that with the single NN, and the results in the figure 
are given for the choice of this time in each case which 
gives the best fit to the full 
evolution\footnote{The values of this time determined in
the way are completely consistent with the model and can 
be understood in greater detail by a closer analysis (see
Ref.~\cite{sl3}).}. The time $t_{\rm max}$ at which the 
comparisons are given in this figure are the latest time
up to which the NN approximation gives a close fit to
the full evolution, i.e., the largest time up to which 
this two phase model of the evolution works well in each 
case.

\subsection{Phase 3: Self-similar evolution}

\begin{figure}
\includegraphics[width=0.5\textwidth]{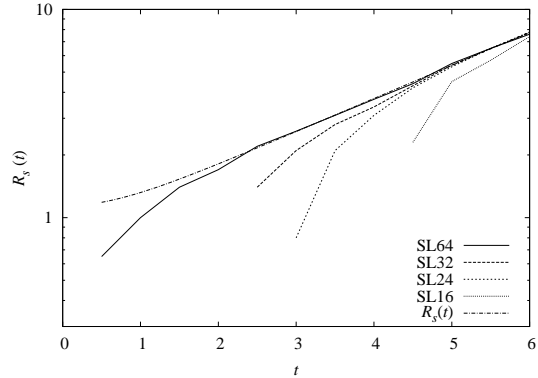}
\caption{Evolution of the rescaling factor $R_s(t)$ in the different
  simulations. Also shown is the self-similar behavior
  Eq.~\eqref{eq:predRst}. From Ref.~\cite{sl1}.
\label{fig:lambdat} }
\end{figure}

We have discussed above how the evolved SL approximates very
well, at sufficiently long times, the ``self-similar'' scaling
corresponding to the temporal behavior of the 
function $R_s(t)$ predicted by the linearized fluid theory.
While above we showed the results for the SL128 simulation, we
have carried out in Ref.~\cite{sl1} the same analysis for
the other initial conditions, and the results are shown in
Fig.~\ref{fig:lambdat}. 
We see that in each case the tendency to join the asymptotic
scaling behaviour is observed, but occurs at a later time. 
This is simply because as the initial $\delta$ decreases
(i.e. the normalised shuffling, cf. Table~\ref{tab:sl_summary})
the duration of the PLT phase increases, and thus the NN
phase is entered later, and breaks down later. Comparing 
the estimated breakdown times $t_{\rm max}$ for the two phase 
approximation given in the caption of Fig.~\ref{fig:linear_NN} 
above with Fig.~\ref{fig:lambdat}, we see 
that each system is close to joining the
self-similar behavior at the time when the NN approximation
breaks down.

We do not have, currently, an analytical or semi-analytical
treatment which allows us to understand the essential 
physical question of the origin of the functional form of 
the {\it spatial dependence of the asymptotic correlation function}. 
However the following
observation, which has been emphasized in Refs. \cite{sl1,sl2,sl3}
is important. Both Figs.~\ref{fig:lambdat} and \ref{fig:Rst} are
derived in the approximation that the correlation function is
fit by a single functional form. To the extent that such a fit
is reasonably good, it means that the correlation function
--- in the relevant range of scale, in which $\xi(r)$ ranges
from close to $10^2$ down to about $0.1$ --- has approximately
the spatial dependence of the asymptotic correlation function.
This suggests strongly --- but does not prove --- that this 
asymptotic form of 
the non-linear correlations is in fact determined by the
early time clustering, in the NN dominated phase, while
its temporal behavior is determined by the fluid limit.
We will discuss this point further in the next section, 
because it is of direct relevance to the question
we consider concerning the role of discreteness in simulations
of this kind.

\section{Some open issues}

We have underlined the qualitative similarity of the evolution
of the infinite SL configurations to that observed in numerical
studies of gravitational clustering in the universe. The
simulations in this context differ from those here 
in the initial perturbations applied to the lattice, and 
the fact that the expansion of the spatial background is 
incorporated [using the equations of motion Eq.~(\ref{Nbody-eom-cosmo})
instead of Eq.~(\ref{Nbody-eom})]. We conclude that the model
we have studied, in which we neglect these extra complexities 
specific to cosmology, provides an interesting ``toy model'' to
address some essential physical questions about this wider
class of simulations. Further we have framed the model in
a purely statistical mechanics language so that these physical
issues, which are related to ones addressed in the wider 
context of long range interacting systems (as in this book), may
be more easily accessible to non-cosmologists.

We thus conclude with a discussion of some specific questions 
which we can address with this ``toy model'' and which
are of direct relevance to the analogous cosmological simulations.
We focus mostly on the issue of discreteness effects, and then
briefly discuss the more general, but related, problem of gaining
a better theoretical understanding of the non-linear regime of
gravitational clustering in these infinite systems.

\subsection{Discreteness effects and the VP limit}

In current cosmological theories of structure formation of the
universe the dominant clustering component is purely self-gravitating
(``dark'') non-relativistic (``cold'') matter. Theoretically it
is described (see e.g. \cite{peebles}) by a set of Vlasov-Poisson(VP) 
equations. Although, as mentioned above, there is no rigorous
derivation demonstrating the applicability of this limit in this 
context, it can be understood as an appropriate (mean-field)
approximation because there is a separation of length scales,
between those characterising the granularity of the
{\it microscopic} physical particles and the {\it astrophysical} 
scales relevant to cosmology. This allows one to view the VP
equations as obtained by a coarse-graining over an intermediate
scale (see e.g. Ref.\cite{buchert_dominguez}). 
 
Solving the VP equations for the $6$-dimensional one particle
phase space density is, however, not feasible numerically, because  
of the strong non-linear coupling of scales in the problem.
For this reason cosmologists adopt the $N$-body
approach, simulating self-gravitating ``macro-particles'' 
in real space. The mass of these unphysical particles are
typically $10^{60-80}$ larger than the supposed mass of the
physical particles, and thus the differences between the
lattice spacing $\ell$ and a physical particle separation
is at least of order $10^{20}$! The problem of determining
the effects of this discretization on the measured 
clustering is the ``discreteness problem''. Until now it
is a problem which has been addressed mostly in a very
qualitative manner, and there is considerable disagreement
in the literature amongst the few authors who have attempted
to address it (see Ref.~\cite{discreteness2_mjbm} for
references). It is now in fact a problem of very practical
relevance as extremely precise predictions will be needed 
in the near future from these simulations in order to compare 
the current theoretical models with the experimental 
results produced by several different observational techniques.

Let us just briefly consider the issue within the context of
the system we have discussed here. Is the evolution of the
system described by the VP equations? While we have shown that
the temporal dependence of the asymptotic self-similar behaviour 
can be derived within this framework, we have seen in the previous
section that we can understand aspects of the evolution using
a NN approximation, which are clearly incompatible with a VP
limit. So what is the VP limit for our system? How do we 
extrapolate to such a limit?

The class of infinite SL we considered is characterised, as we
discussed by just two parameters, which we can choose to be
$\ell$ and $\delta$. Gravity, however, is a scale free
and so, dynamically, there is really only one parameter: any
two SL with the same $\delta$ are in fact equivalent up to
a redefinition of the unit of length (which may be chosen
equal to $\ell$). Thus $\delta$ is in fact the single 
physically relevant parameter. Can we recover a VP
approximation for some limiting value of $\delta$?
The answer is clearly in the negative: for any value
of $\delta$ the NN phase remains\footnote{We have discussed only
explicitly the case $\delta < 1$. For any $\delta >1$ the SL
at small scales is identical to a Poisson distribution. In this
case the NN phase occurs always at the beginning of the evolution.
For more detail, see Refs.~\cite{Baertschiger:2004tx, sl3}.}.
It is interesting to note also that as $\delta \rightarrow 0$ the PLT treatment
we described becomes valid for an arbitrarily long time.
In fact it can be shown\cite{marcos_06, discreteness2_mjbm} 
that this leads to a divergence of the evolution from the VP 
limit, given in this case by the linearized fluid evolution. 
In fact PLT approximates well the fluid evolution on time 
scales of a few dynamical times, but deviates more and 
more in time, with an asymptotic divergence\footnote{We note that 
this finding makes sense physically as the VP limit is taken at fixed
time. We consider here instead the limit in which the particle density
is held fixed but the time diverges.}.

To define a VP limit for our infinite system we clearly need a length 
scale. Indeed as we noted in our discussion of the comparison of the
thermodynamic limit and the VP limit above, the VP limit
must be one in which the particle density $n_0 \rightarrow \infty$, 
which corresponds to sending $\ell \rightarrow 0$. We must therefore 
have another length scale to compare the density with. The scale 
which can evidently be used is the smoothing length $\varepsilon$,
in which case the VP limit is $\ell \rightarrow 0$ at fixed 
$\varepsilon$ (and then, if the limit is indeed defined, 
$\varepsilon \rightarrow 0$). Studying what we have learnt about 
the evolution of the SL, we can verify that this limit does
indeed lead to the recovery of the fluid limit of the 
two-phase model we have discussed. Firstly PLT, modified
to include the smoothing, can be shown\cite{marcos_06,
discreteness2_mjbm} to converge, if $\ell \ll \epsilon$,
to the fluid limit. Further the NN dominated phase we have
described disappears also once we reach the regime 
$\ell \ll \epsilon$, as the force due to nearby particles 
is now no longer felt. 

The conclusion of this analysis is that to attain strictly
the VP limit of such simulations we should work in the
parameter range $\ell \ll \varepsilon$. This is {\it not} 
the current practice in cosmological 
simulations(see e.g. Ref.~\cite{springel_05}), but does correspond
to the conclusions of one of the few controlled numerical 
studies of the issue in the literature\cite{splinter}.
While the results of (most) simulations, which use 
$\ell \ll \varepsilon$, are not necessarily grossly
wrong, we conclude that numerical extrapolations
to the regime $\ell > \varepsilon$ will be necessary
to determine how precise they are. 

\subsection{Non-linear structures and quasi-stationary states}

We conclude with a brief discussion of the more general 
question of the nature of non-linear clustering in these systems.
In recent years, in parallel with the description of the results
of simulations in terms of the two-point correlation properties,
cosmologists have developed a different phenomenological description
of the results of simulations, using 
so-called ``halo models'' (see e.g. \cite{halo}).
These models following naturally from the ``hierarchical'' picture
of clustering, in which it is initial small overdensities at a 
given scale which evolve linearly until they collapse, essentially
then behaving like independent finite sub-systems. The non-linear 
structure in simulations can be well described as a collection of
such ``halos'', which are roughly spherical and approximately
virialised structures, with properties very similar to those 
of structures formed after the ``violent relaxation''  of a 
finite gravitating system in an initially unstable configuration.
These ``halos'' are believed to be understood again within
the framework of VP equations, and are thus closely analogous
to the ``quasi-stationary states'' (QSS) which have recently been much
studied in certain statistical mechanical toy models (see contribution
of S. Ruffo in this volume). While they have been characterised in
great detail in huge numerical studies, and in particular certain
apparently ``universal'' properties 
identified (see e.g. Refs.~\cite{navarro2,taylor+navarro,hansen_etal,hansen+moore}) their 
physics is not at all understood. In this respect it is 
interesting to note that a recent detailed study\cite{arad}
concludes that the framework of 
Lynden-Bell, exploited recently in the study of such QSS in simple 
systems, is apparently not useful in predicting their
properties.

In the SL system it appears visually, and would be expected from 
the qualitative features we have seen, that 
such a halo model description of the clustering could be given.
It would be interesting to analyse in this context the nature of
these halos, and also their relation to virialized structures
formed in finite open systems (e.g. in spherical cold collapse).
Further it would be perhaps useful to explore in particular
lower dimensional (i.e. two or one) dimensional models which
resemble more closely the infinite system we have discussed
in order to simplify and possibly make closer contact with the
insights gained, notably about QSS, in the studies of simple 
statistical mechanical models.

I am indebted to my collaborators in the research reported here:
Thierry Baertschiger, Andrea Gabrielli, Bruno Marcos and Francecso
Sylos Labini. I also thank Michael Kiessling and Francois Sicard
for several conversations which were very useful in relation to the
discussion of the first section of this paper.


\begin{thebibliography}{39}
\expandafter\ifx\csname natexlab\endcsname\relax\def\natexlab#1{#1}\fi
\providecommand{\enquote}[1]{``#1''}
\expandafter\ifx\csname url\endcsname\relax
  \def\url#1{\texttt{#1}}\fi
\expandafter\ifx\csname urlprefix\endcsname\relax\def\urlprefix{URL }\fi
\providecommand{\eprint}[2][]{\url{#2}}

\bibitem[Baertschiger et~al.(2007{\natexlab{a}})]{sl1}
T.~Baertschiger, M.~Joyce, A.~Gabrielli, and F.~Sylos~Labini, \emph{Phys. Rev.}
  \textbf{E75}, 021113 (2007{\natexlab{a}}), \eprint{cond-mat/0607396}.

\bibitem[Baertschiger et~al.(2007{\natexlab{b}})]{sl2}
T.~Baertschiger, M.~Joyce, A.~Gabrielli, and F.~Sylos~Labini, \emph{Phys. Rev.}
  \textbf{E76}, 011116 (2007{\natexlab{b}}), \eprint{cond-mat/0612594}.

\bibitem[Baertschiger et~al.(2007{\natexlab{c}})]{sl3}
T.~Baertschiger, M.~Joyce, F.~Sylos~Labini, and B.~Marcos, 
\eprint{arXiv:0711.2219}.

\bibitem[Jeans(1902)]{jeans}
J.~H. Jeans, \emph{Phil. Trans. Roy. Soc.} \textbf{199}, 1--53 (1902).

\bibitem[Binney and Tremaine(1994)]{binney}
J.~Binney, and S.~Tremaine, \emph{Galactic Dynamics}, Princeton University
  Press, 1994.

\bibitem[Kiessling(2003)]{kiessling}
M.~K.-H. Kiessling, \emph{Adv. Appl. Math.} \textbf{31}, 132--149 (2003),
  \eprint{astro-ph/9910247}.

\bibitem[Gabrielli et~al.(2004)]{book}
A.~Gabrielli, F.~Sylos~Labini, M.~Joyce, and L.~Pietronero, \emph{Statistical
  Physics for Cosmic Structures}, Springer, 2004.

\bibitem[Chandrasekhar(1943)]{chandra43}
S.~Chandrasekhar, \emph{Rev. Mod. Phys.} \textbf{15}, 1 (1943).

\bibitem[Gabrielli et~al.(2006)]{gabrielli_06}
A.~Gabrielli, T.~Baertschiger, M.~Joyce, B.~Marcos, and F.~Sylos~Labini,
  \emph{Phys. Rev.} \textbf{E74}, 021110 (2006), \eprint{cond-mat/0603124}.

\bibitem[Gabrielli(200X)]{andrea_1dforces}
A.~Gabrielli, \emph{Phys. Rev.} \textbf{E72}, 066113 (2005),
  \eprint{cond-mat/0506365}.

\bibitem[Hernquist et~al.(1991)]{hernquist}
L.~Hernquist, F.~R. Bouchet, and Y.~Suto, \emph{Astrophys. J.} \textbf{75},
  231--240 (1991).

\bibitem[Efstathiou et~al.(1985)]{efstathiou_init}
G.~Efstathiou, M.~Davis, C.~S. Frenk, and S.~D.~M. White, \emph{Astrophys. J.
  Supp.} \textbf{57}, 241--260 (1985).

\bibitem[Peebles(1980)]{peebles}
P.~J.~E. Peebles, \emph{{The Large-Scale Structure of the Universe}}, Princeton
  University Press, 1980.

\bibitem[Padmanabhan(1990)]{Padmanabhan:1989gm}
T.~Padmanabhan, \emph{Phys. Rept.} \textbf{188}, 285 (1990).

\bibitem[Spohn(1991)]{spohn}
H.~Spohn, \emph{Large Scale Dynamics of Interacting Particles},
  Springer-Verlag, 1991.

\bibitem[Braun and Hepp(1977)]{braun+hepp}
W.~Braun, and K.~Hepp, \emph{Comm. Math. Phys.} \textbf{56}, 101--113 (1977).

\bibitem[Ruelle(1983)]{ruelle}
D.~Ruelle, \emph{{Statistical Mechanics: Rigorous results}}, W. A. Benjamin,
  1983.

\bibitem[Gabrielli et~al.(2002)]{glasslike}
A.~Gabrielli, M.~Joyce, and F.~Sylos~Labini, \emph{Phys. Rev.} \textbf{D 65},
  083523 (2002).

\bibitem[Torquato and Stillinger(2003)]{to03}
S.~Torquato, and F.~H. Stillinger, \emph{Phys. Rev.} \textbf{E68}, 041113
  (2003), publisher's note: Phys. Rev. E, \textbf{68}, 069901 (2003).

\bibitem[Joyce and Marcos(2007)]{discreteness1_mjbm}
M.~Joyce, and B.~Marcos, \emph{Phys. Rev.} \textbf{D75}, 063516 (2007),
  \eprint{astro-ph/0410451}.

\bibitem[{\textnormal{\texttt{www.mpa-garching.mpg.de/gadget/right.html}}}(200%
0)]{gadget}
{\textnormal{\texttt{www.mpa-garching.mpg.de/gadget/right.html}}} (2000).

\bibitem[Springel et~al.(2001)]{gadget_paper}
V.~Springel, N.~Yoshida, and S.~D.~M. White, \emph{New Astronomy} \textbf{6},
  79--117 (2001), (also available on \cite{gadget}).

\bibitem[{Efstathiou} et~al.(1988)]{efstathiou_88}
G.~{Efstathiou}, C.~S. {Frenk}, S.~D.~M. {White}, and M.~{Davis}, \emph{Mon.
  Not. R. Astron. Soc.} \textbf{235}, 715--748 (1988).

\bibitem[Jain and Bertschinger(1996)]{bertschinger}
B.~Jain, and E.~Bertschinger, \emph{Astrophys. J.} \textbf{456}, 43--54 (1996).

\bibitem[Jain and Bertschinger(1998)]{bertschinger2}
B.~Jain, and E.~Bertschinger, \emph{Astrophys. J.} \textbf{509}, 517--530
  (1998).

\bibitem[Cooray and Sheth(2002)]{halo}
A.~Cooray, and R.~Sheth, \emph{Phys. Rep.} \textbf{379}, 1--129 (2002).

\bibitem[Peacock and Dodds(1996)]{peacock}
J.~A. Peacock, and S.~J. Dodds, \emph{Mon. Not. R. Astron. Soc.} \textbf{280},
  L19--L26 (1996).

\bibitem[Pines(1963)]{pines}
D.~Pines, \emph{Elementary Excitations in Solids}, Benjamin, New York, 1963.

\bibitem[Marcos et~al.(2006)]{marcos_06}
B.~Marcos, T.~Baertschiger, M.~Joyce, A.~Gabrielli, and F.~Sylos~Labini,
  \emph{Phys. Rev} \textbf{D73}, 103507 (2006), \eprint{astro-ph/0601479}.

\bibitem[Buchert and Dominguez(2005)]{buchert_dominguez}
T.~Buchert, and A.~Dominguez, \emph{Astron. Astrophys.} \textbf{438}, 443--460
  (2005).

\bibitem[Joyce and Marcos(2006)]{discreteness2_mjbm}
M.~Joyce, and B.~Marcos, \emph{Phys. Rev.} \textbf{D76}, 103505 (2007),
  \eprint{arXiv:0704.3697}.

\bibitem[Baertschiger and Sylos~Labini(2004)]{Baertschiger:2004tx}
T.~Baertschiger, and F.~Sylos~Labini, \emph{Phys. Rev.} \textbf{D69}, 123001
  (2004), \eprint{astro-ph/0401238}.

\bibitem[Springel et~al.(2005)]{springel_05}
V.~Springel, et~al., \emph{Nature} \textbf{435}, 629--636 (2005),
  \eprint{astro-ph/0504097}.

\bibitem[Splinter et~al.(1998)]{splinter}
R.~J. Splinter, A.~L. Melott, S.~F. Shandarin, and Y.~Suto, \emph{Astrophys.
  J.} \textbf{497}, 38--61 (1998).

\bibitem[Navarro et~al.(1997)]{navarro2}
J.~F. Navarro, C.~S. Frenk, and S.~D.~M. White, \emph{Astrophys. J.}
  \textbf{490}, 493--508 (1997).

\bibitem[Taylor and Navarro(2001)]{taylor+navarro}
J.~Taylor, and J.~Navarro, \emph{Astrophys. J.} \textbf{563}, 483 (2001).

\bibitem[Hansen et~al.(2006)]{hansen_etal}
S.~Hansen, B.~Moore, M.~Zemp, and J.~Stadel, \emph{JCAP} \textbf{0601}, 014
  (2006).

\bibitem[Hansen and Moore(2006)]{hansen+moore}
S.~Hansen, and B.~Moore, \emph{New Astronomy} \textbf{11}, 333 (2006).

\bibitem[Arad and Johansson(2005)]{arad}
I.~Arad, and P.~Johansson, \emph{Mon. Not. R. Astron. Soc.} \textbf{362}, 252
  (2005).

\end{thebibliography}


\end{document}